\documentclass[a4paper,11pt]{article}
\pdfoutput=1 % if your are submitting a pdflatex (i.e. if you have
\usepackage{jheppub} % for details on the use of the package, please
\usepackage[T1]{fontenc} % if needed
\usepackage{caption} 
\usepackage{subcaption}

\usepackage[latin9]{inputenc}
\usepackage{float}
\usepackage{amsmath}
\usepackage{amssymb}
\usepackage{graphicx}
\usepackage{esint}
 \usepackage{hyperref}
\usepackage{comment}
\usepackage{color}
\usepackage{microtype}
\usepackage{cleveref}
\usepackage{breakurl}
\usepackage{bbm}
\newcommand{\be}{\begin{equation}}
\newcommand{\ee}{\end{equation}}
\newcommand{\ben}{\begin{displaymath}}
\newcommand{\een}{\end{displaymath}}
\newcommand{\bea}{\begin{eqnarray}}
\newcommand{\eea}{\end{eqnarray}}
\def\K{K{\"a}hler}
   \newcommand{\rf}[1]{(\ref{#1})}

\def\be{\begin{equation}}
\def\ee{\end{equation}}
\def\bea{\begin{eqnarray}}
\def\eea{\end{eqnarray}}
\def\ba{\begin{array}}
\def\ea{\end{array}}
\def\bit{\begin{itemize}}
\def\eit{\end{itemize}}

 \makeatletter

\allowdisplaybreaks

\makeatother

\makeatletter
\DeclareRobustCommand{\rcite}[1]{%
  \rcite@aux#1,\@nil{#1}%
}
\def\rcite@aux#1,#2\@nil#3{%
  \if\relax#2\relax
    Ref.~\cite{#3}%
  \else
    Refs.~\cite{#3}%
  \fi
}
\makeatother

\hypersetup{
    colorlinks = true,
    citecolor = {blue},
    linkcolor = {blue},
    urlcolor = {blue},
}

 \title{\rm {\bf \huge  \boldmath Planck 2018 and Brane Inflation Revisited}}

\author{Renata Kallosh,}
\author{Andrei Linde,}
\author{and Yusuke Yamada}
\affiliation{Stanford Institute for Theoretical Physics and Department of Physics, Stanford University, Stanford, CA 94305, USA}
\emailAdd{kallosh@stanford.edu}
 \emailAdd{alinde@stanford.edu}
  \emailAdd{yusukeyy@stanford.edu}

\notoc

\abstract{We revisit phenomenological as well as string-theoretical aspects of D-brane inflation cosmological models. Phenomenologically these models  stand out on  par with $\alpha$-attractors, as models with Planck-compatible values of $n_s$, moving  down to the sweet spot in the data with decreasing value of $r$.  On the formal side we present a new supersymmetric version of these models in the context of de Sitter supergravity with a nilpotent multiplet and volume modulus stabilization. The geometry of the nilpotent multiplet is evaluated in the framework of string theory.}

\begin{document}

\maketitle

 \newpage 
  \tableofcontents{}

 \parskip 6pt 
\section{Introduction}
\begin{figure}[!h]
%\vspace*{3mm}
%\hspace{-3mm}
\begin{center}
\hskip 0.3cm\includegraphics[width=15cm]{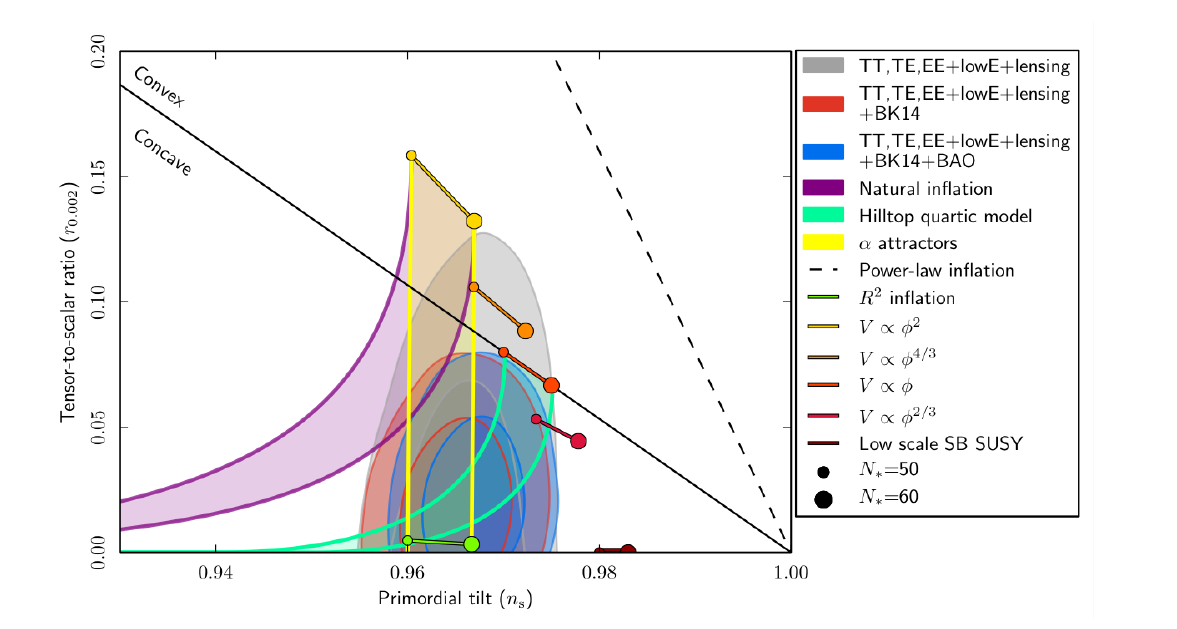}
\caption{\footnotesize The results of the Planck 2018 data release. The two yellow lines correspond to $\alpha$-attractors with
$n_s \approx 1-{2\over N_e}$, with the number of e-foldings $N_{e} = 50$ and $N_{e} = 60$.}
\label{Planck}
\end{center}
\vspace{0cm}
\end{figure}

D-brane inflation models have two aspects, a phenomenological and a  string-theoretical. Both became very interesting after the investigation of inflationary models in the Planck 2018 data release~\cite{Akrami:2018odb}.   The Planck 2018 $n_s$~-~$r$ plane is shown in Fig.~\ref{Planck}. The dark (light) blue regions describe the $1\sigma$ ($2\sigma$) confidence level for the CMB related  data obtained by Planck 2018 and  Bicep/Keck2014, additionally including the baryon oscillations (BAO) data. Meanwhile the comparison of  predictions of inflationary models with the data in~\cite{Akrami:2018odb} was based on the CMB related  data only, excluding BAO. The corresponding 1$\sigma$ and 2$\sigma$ regions are shown in red in  Fig.~\ref{Planck}.

The main part of the left hand side of the $1\sigma$ region in this plane is described by the simplest $\alpha$-attractor model~\cite{Kallosh:2013yoa}, with the predictions bounded by the two yellow lines  corresponding to the $\alpha$-attractor prediction 
\be\label{alphans}
n_{s} = 1-{2\over N_{e}} \, 
\ee
with the number of e-foldings $N_{e} = 50$ and $N_{e} = 60$. The lower part of the $\alpha$-attractor yellow band covers also the predictions of the Starobinsky model ~\cite{Starobinsky:1980te}, the GL supergravity model~\cite{Goncharov:1985yu}, and the Higgs inflation model~\cite{Salopek:1988qh,Bezrukov:2007ep}. The values of $n_s$ and $r$ for $\alpha$-attractor models  were shown in~\cite{Kallosh:2013yoa} and~\cite{Carrasco:2015pla}, see Fig.~\ref{alpha} here. Special cases with $\alpha=1$ correspond to Starobinsky, Higgs and conformal inflation~\cite{Kallosh:2013hoa},  $\alpha = 1/9$ corresponds to the GL model~\cite{Goncharov:1985yu}, $\alpha=2, 1/2$ correspond to fibre inflation~\cite{Cicoli:2008gp,Kallosh:2017wku}. Models with $3\alpha=7,6,5,4,3,2,1$ originate from theories with maximal supersymmetry~\cite{Ferrara:2016fwe,Kallosh:2017ced}.

\begin{figure}[!t]2
%\vspace*{3mm}
%\hspace{-3mm}
\begin{center}
\includegraphics[scale=0.35]{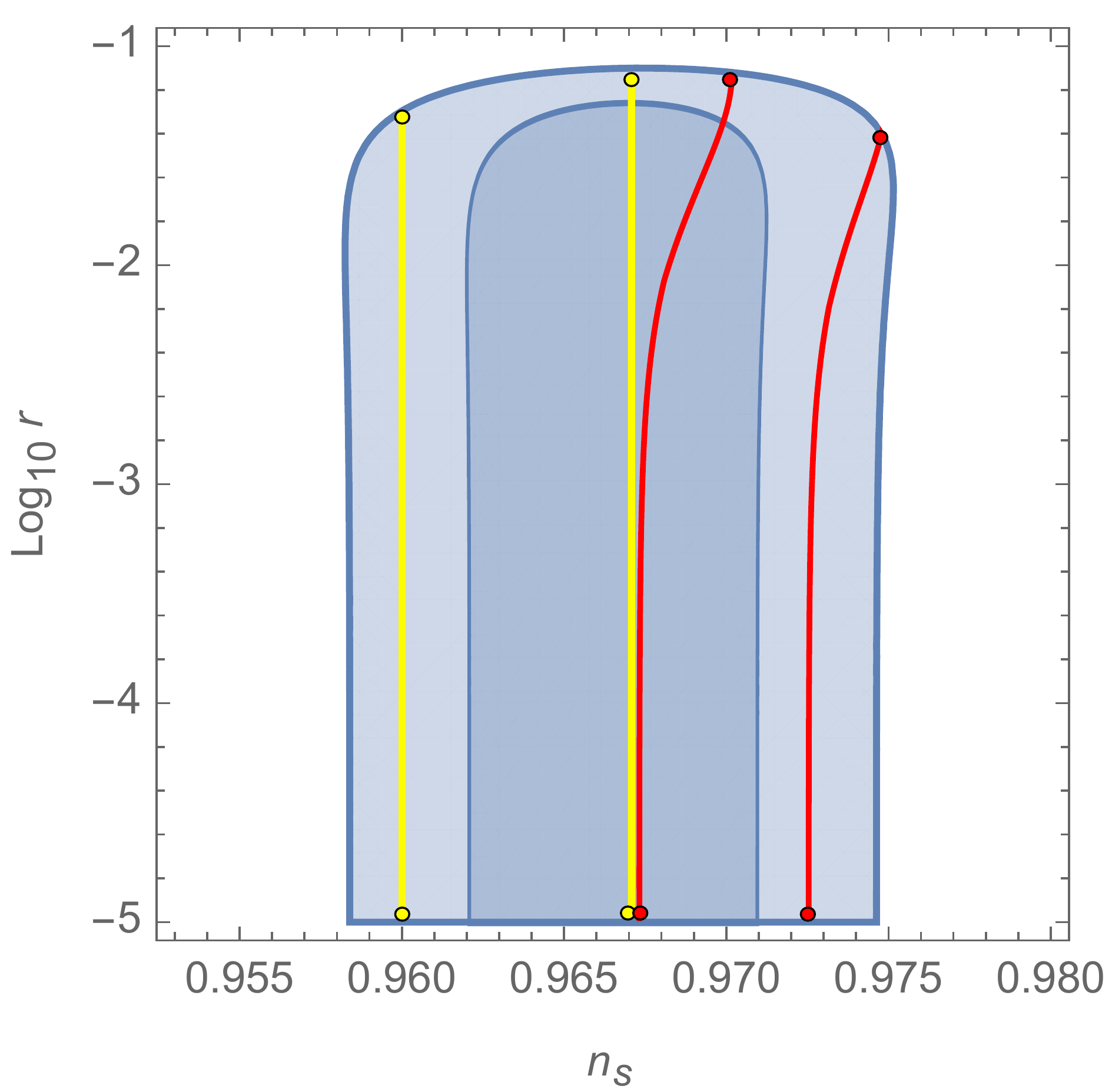} \hskip 15pt
\includegraphics[scale=0.35]{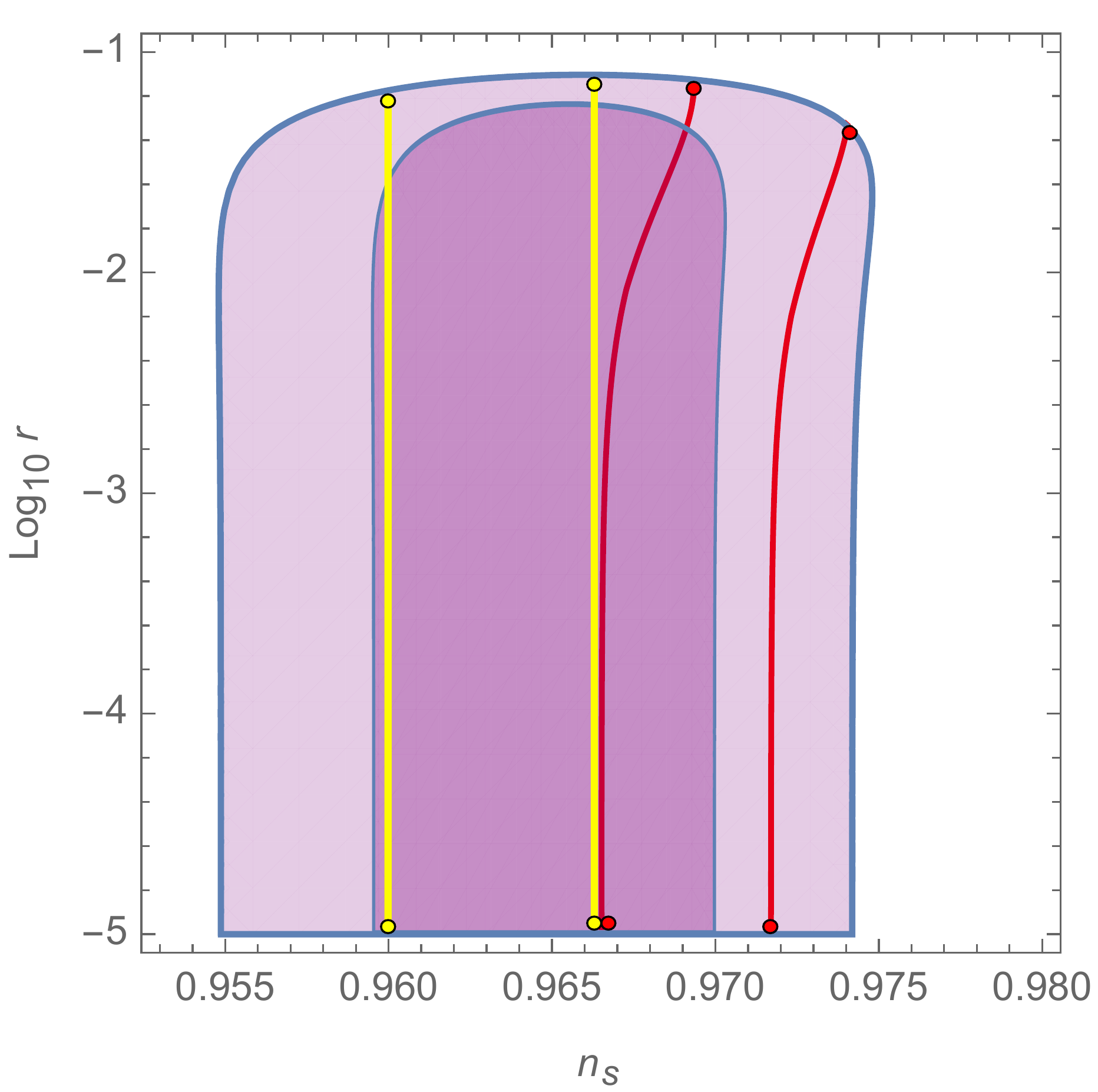}
\end{center}
\vspace{-0.5cm}
\caption{\footnotesize Comparison of predictions of $\alpha$-attractors  and  of the D-brane inflationary model with $V \sim 1- \Big ({m\over \phi}\Big )^4$ within the $2\sigma$ area of  the Planck 2018 results for $n_{s}$ and $r$. On the left panel, the dark (light) blue area is the Planck 2018 $1\sigma$  ($2\sigma$) region, with an account taken of the CMB-related data.  The right panel represents the Planck 2018 results based on the CMB related data only, without  BAO. Two yellow lines on both panels are for the quadratic T-model of $\alpha$-attractors at $N_e = 50$ and $N_e = 60$.  Two red lines are for the simplest D-brane inflation model $V =1- \Big ({m\over \phi}\Big )^4$.}
\label{Quartic}
\end{figure}

As one can see from~\cite{Akrami:2018odb}, the two yellow lines  corresponding to $\alpha$-attractors cover  {\it the left hand side} of the $1\sigma$ dark blue (and dark red) areas in Fig.~\ref{Planck}.
However, the Planck 2018 analysis of  inflationary models presented in  Table 5 in~\cite{Akrami:2018odb} reveals yet another class of models, which may match the observations equally well, in a complementary way.  It is a class of D-brane inflation models with the potential proportional to  $1- \Big ({m\over \phi}\Big )^4 +\cdots $,   studied in  Appendix C of the KKLMMT paper~\cite{Kachru:2003sx}. Predictions of such models cover {\it the right hand side} of the $1\sigma$ dark blue (and dark red) areas in Fig.~\ref{Planck}.

As an example, let us simultaneously plot the predictions of $\alpha$-attractors and of the simplest  D-brane inflationary model with $V \sim 1- \Big ({m\over \phi}\Big )^4$ in figures representing the Planck 2018 data for $r$ on $\log_{10}r$ scale, which is more suitable for illustration of the predictions of the models in the limit of small $r$, where both of these classes of models exhibit attractor behavior, see Fig.~\ref{Quartic}. As one can see from this figure, the combination of the simplest $\alpha$-attractor model  and the simplest D-brane inflation model almost completely cover the $1\sigma$ dark blue (dark purple) area of the Planck 2018 data. 

The reason why the predictions of $\alpha$-attractors  and of the simplest D-brane inflation  match each other  in Fig.~\ref{Quartic} so perfectly is very simple. According to the investigation in  Appendix C of~\cite{Kachru:2003sx},  the value of $n_{s}$ in the D-brane inflationary model with $V \sim 1- \Big ({m\over \phi}\Big )^4$ in the small $m$ (small $r$) limit is given by
\be\label{apC}
n_s = 1-{5\over 3 N_{e}} \ . 
\ee
The cosmological evolution in these models was studied in detail in~\cite{Lorenz:2007ze}  and in Sect. 5.19 of~\cite {Martin:2013tda}, see Figs.~159-166 there.

Magically, $n_{s}$ for $\alpha$-attractors shown by the right yellow line corresponding to $N_{e} = 60$ in  Fig.~\ref{Quartic} exactly coincides with $n_{s}$ for the simplest D-brane inflation models for the left red line with $N_{e} = 50$:
\be
n_{s}=1-{2\over 60} = 1-{5\over 3\times 50} = 1-{1\over 30} = 0.967 \ .
\ee
This explains why $\alpha$-attractors and D-brane inflation match Planck results  so well in Fig.~\ref{Quartic}, and why both models provide a very good match to the  Planck 2018 result $n_s = 0.9649 \pm  0.0042$  \cite{Akrami:2018odb}. 

Of course, one should remember that exact predictions of these models depend on details of the models, mechanism of reheating, etc., and the position of the $1\sigma$ region for Planck 2018 depends on the data set (e.g. with or without BAO). Nevertheless the perfect match shown in Fig.~\ref{Quartic} is quite striking.

Note, that the predictions of $\alpha$-attractors also allow some variability, converging to \rf{alphans} in the limit of large $N_{e}$, and small $\alpha$, which corresponds to small $r$, see Fig.~\ref{alpha}.
\begin{figure}[h!]
%\vspace*{3mm}
%\hspace{-3mm}
\begin{center}
 \includegraphics[width=10cm]{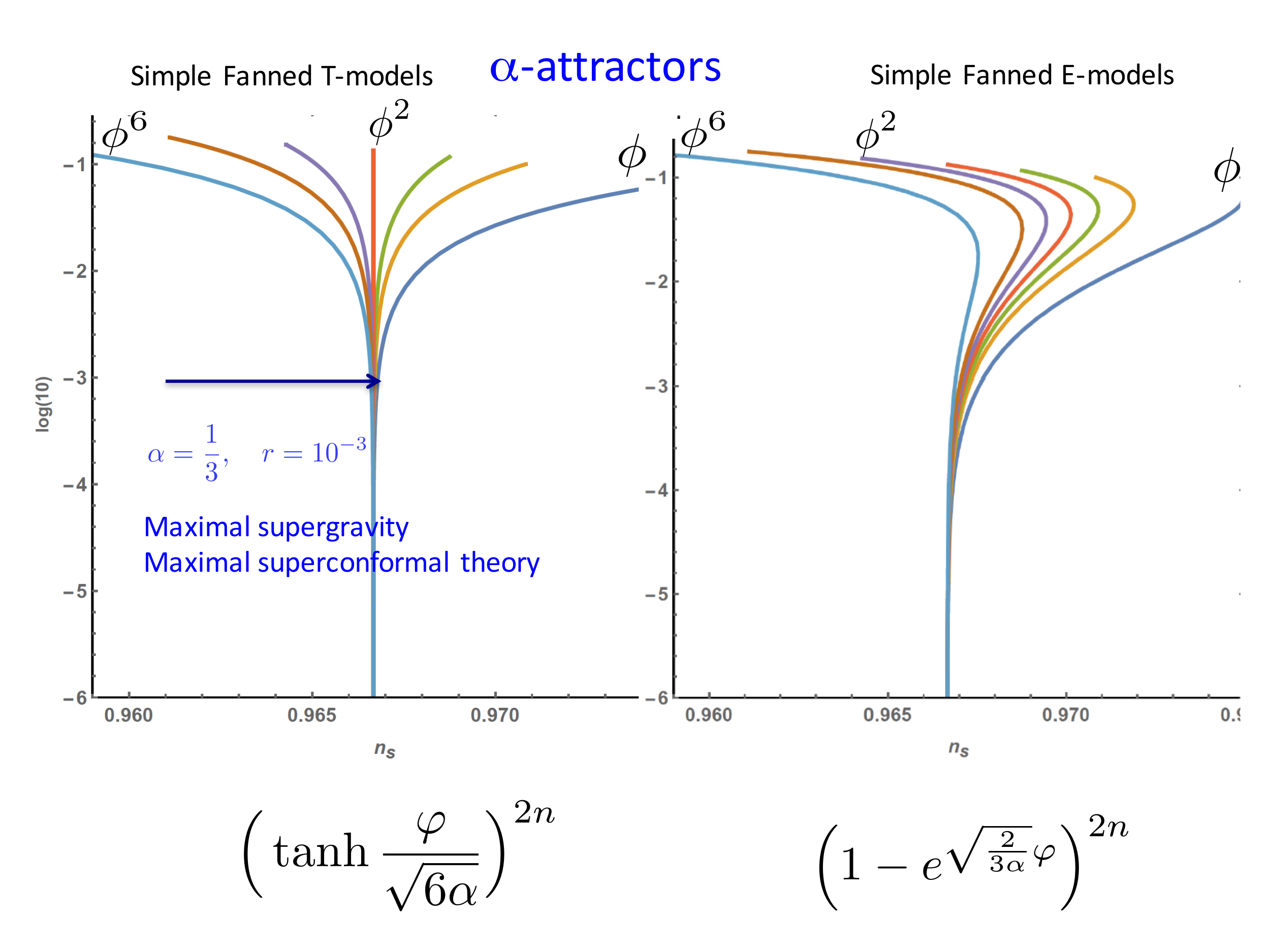}
\end{center}
\vspace*{-6mm}
\caption{\footnotesize $\alpha$-attractor models in $n_s$~-~$r$ plane at $N_e=60$.}
\label{alpha}
\vspace{0cm}
\end{figure}

Thus, phenomenologically D-brane inflation model after Planck 2018 has acquired a new significance, even independent of its string theory origin. But we will show here that a new progress can be made  with regard to string theory implementation of the phenomenologically attractive versions of D-brane inflation model.

The string theory origin of D-brane inflation model is often attributed to KKLMMT model~\cite{Kachru:2003sx}, where D3-brane-$\overline {\rm D3}$-brane interaction was studied in the context of the volume modulus stabilization. Earlier proposals for D-brane inflation relevant to our current discussion  were made in~\cite{Dvali:2001fw,Burgess:2001fx,GarciaBellido:2001ky}.  The  inflationary potentials  corresponding to Dp-brane-$\overline {\rm Dp}$-brane interaction were proposed in the form
\be
V_{BI}= V_{0}\Big ( 1-\Big({m\over \phi}\Big) ^{7-p} +\cdots \Big)\, ,  \qquad V_{KKLTI}= V_{0} \Big ( 1+\Big({m\over \phi}\Big) ^{7-p} \Big)^{-1} .
\label{BI}\ee
In \cite {Martin:2013tda} a detailed cosmological analysis of both potentials was performed, with emphasis on the  cases  $p=3$ and $p=5$, i.e. for  inverse quartic and quadratic potentials. 
The first potential in \rf{BI} was called BI (Brane Inflation), the second one was called  KKLTI (KKLT Inflation).

The earlier proposals in~\cite{Dvali:2001fw,Burgess:2001fx,GarciaBellido:2001ky} were made without addressing the volume stabilization issue. A consistent cosmological evolution of Dp-branes in string theory with the fundamental ten-dimensional geometry can be interpreted as an evolution in the   four-dimensional Einstein equations under condition that 
 the six-dimensional internal space has a constant volume,  not a runaway behavior which destroys the consistency of the four-dimensional cosmology.

 In~\cite{Kachru:2003sx} D3-brane-$\overline{\rm D3}$-brane interaction was studied simultaneously with the volume stabilization. The concrete example in string theory was based on strong warping, which is dual to an almost conformal four-dimensional field theory. Therefore the scalar field describing the motion of the brane was conformally coupled  to gravity. 
 This was consistent with the  choice of the \K\, potential and the KKLT superpotential
 in the form
\be
K=-3\log(T+  \bar T - \Phi \Bar \Phi)\, ,\qquad 
W= W_0 +A(\Phi) e^{-T} .
\label{KWDym}\ee
However, it is well known that it is difficult to achieve inflation in the theories with a conformally coupled inflaton.  This was one of the the main reasons to consider a generalized KKLT superpotential \rf{KWDym} where the  dependence on the distance between branes $\Phi$ was included in $A(\Phi)$. A specific explicit model of inflation which followed from~\rf{KWDym} was presented in~\cite{Baumann:2007ah}. The corresponding potential, an inflection point potential, has $n_s\approx 0.93$, which is ruled out by the data, see for example Fig.~\ref{Planck}. More about the derivation of this model can be found in a review paper on string cosmology~\cite{Baumann:2014nda}.

The reason for us to revisit the KKLMMT paper~\cite{Kachru:2003sx} after Planck 2018  is that, in addition and independently of an example of a  particular form of the \K\,  potential and superpotential~\rf{KWDym},  Ref.~\cite{Kachru:2003sx} provided a basis for other, more general approaches to string cosmology.  It can be used at present in a form
 which is in fact supported by the latest data, using in particular equation~\rf{apC}  for the tilt of the spectrum as derived in Appendix C of~\cite{Kachru:2003sx}. We will therefore start with phenomenological properties of D-brane inflation models following~\cite{Kachru:2003sx,Lorenz:2007ze, Martin:2013tda},  and then we will discuss a possibility to implement such models in supergravity and string theory. 

\section{Phenomenological D-brane inflation models}

\subsection{KKLMMT scenario and inverse quartic and quadratic models}

\begin{figure}[t!]
\begin{center}
\includegraphics[scale=0.35]{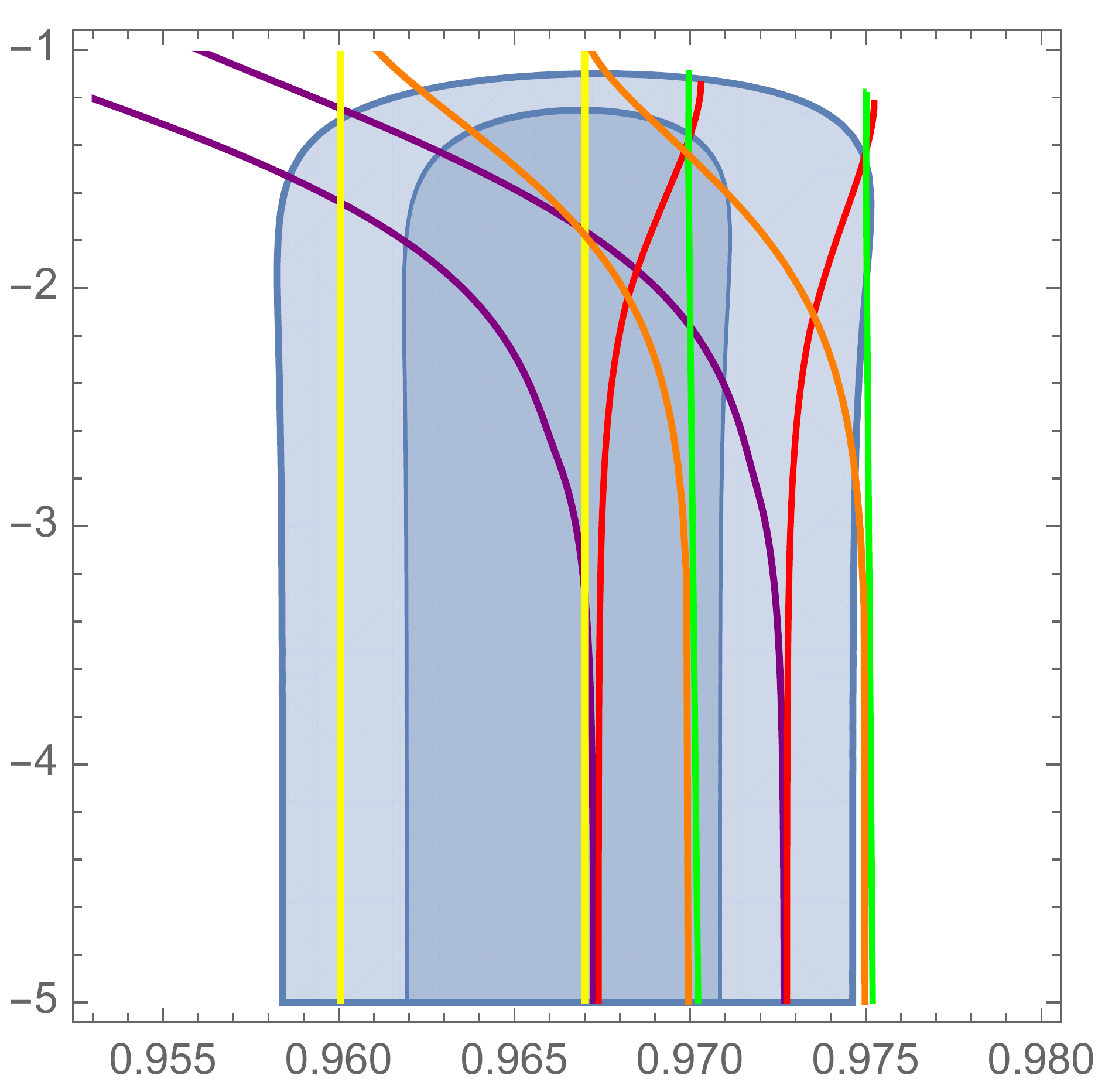}
\end{center}
\caption{\footnotesize \footnotesize Predictions of $\alpha$-attractors  and KKLTI/BI models. The dark (light) blue area is the Planck 2018 $1\sigma$  ($2\sigma$) region, with an account taken of TT, TE, EE + lowE + lensing + BK14+BAO~\cite{Akrami:2018odb}.  Two yellow lines on the left panel are for the quadratic T-model of $\alpha$-attractors at $N_e = 50$ and $N_e = 60$.  Two purple lines are for the quartic KKLTI model, two red lines are for the quartic BI model. As one can see, the combination of the simplest $\alpha$-attractor model and the quartic KKLTI/BI model (including the models interpolating between KKLTI and BI) completely cover the $1\sigma$ area of the Planck 2018 data. Two orange lines  show the predictions of the quadratic KKLTI model, and two green lines are for the quadratic BI model. }
\label{Quartic1}
\end{figure}

A short preview of  the phenomenology of the simplest D-brane  inflationary models  is given in  Figure~\ref{Quartic}. In this section we will study a broad class of phenomenological models of D-brane  inflation in greater detail. As in Encyclopedia Inflationaris~\cite {Martin:2013tda}, we will call the corresponding potentials either BI (Brane Inflation) for a Coulomb-type interaction, or 
KKLTI (for KKLT Inflation) when the potential takes a form of the inverse harmonic function. 
The case of the  inverse quadratic Coulomb-type interaction is
\be
^2 V_{BI}=V_0 \Big [1- \Big ({m\over \phi}\Big )^2 +\cdots  \Big ]\, , 
\label{V2}\ee
whereas the potential in a form of the inverse harmonic function is
\be
 ^2V_{KKLTI}=V_0 \Big [1+ \Big ({m\over \phi}\Big )^2 \Big ]^{-1} = V_0 \Big [{\phi^2\over 
m^2 + \phi^2} \Big ] \ .
\label{V2exact}\ee
At small ${m\over \phi}$ the `exact' non-singular at $\phi=0$ potential $V_{KKLTI}$ takes the form of $V_{BI}$. At very large values of $m$ the potential tends to a quadratic one, in the area where  $\phi  \ll m$,
\be
 ^2V_{KKLTI}\Rightarrow {V_0\over m ^2} \phi^2.
\label{p2}\ee
In the limit $m\gg 1$, predictions of this model coincide with the predictions of the simplest chaotic inflation model $V \sim \phi^{2}$, which is ruled out. However, as we will see soon, in the case $m \lesssim 1$ this model provides a good fit to Planck 2018 data, see Fig. \ref{Quartic1}.

The case of the inverse quartic Coulomb-type interaction is
\be
^4V_{BI}=V_0 \Big [1- \Big ({m\over \phi}\Big )^4 +\cdots  \Big ]\, ,  
\label{V4}\ee
whereas the potential in a form of the inverse harmonic function is
\be
 ^4V_{KKLTI}=V_0 \Big [1+ \Big ({m\over \phi}\Big )^4 \Big ]^{-1} =V_0 \Big [{\phi^4\over 
m^4 + \phi^4} \Big ] \ .
\label{V4exact}\ee
Here again, at small ${m\over \phi}$ the `exact' non-singular at $\phi=0$ potential $V_{KKLTI}$ takes the form of $V_{BI}$. At very large values of $m$ the potential tends to a quartic one, in the area where  $\phi  \ll m$,
\be
 ^4V_{KKLTI} \Rightarrow {V_0\over m^4} \phi^4.
\label{p4}\ee

\noindent  The $\alpha$-attractor models have 
\be
 n_s \approx  1- {2 \over  N_e}.
\ee
For Dp-brane-$\overline {\rm Dp}$-brane inflation models with $V= A-{B\over \phi^{7-p}}$,  the general formula for small $r$ is
\be
 n_s \approx  1- {2 (8-p) \over  (9-p)N_e}  .
\label{gen}
\ee

This  was also given in~\cite{Burgess:2001fx} and in~\cite {Martin:2013tda} in slightly different notation. For example in~\cite {Martin:2013tda} the formula is in terms of $k= 7-p$ and is given as $ n_s \approx  1- {2 (k+1) \over  (k+2)N_e}  $.
Note that for the polynomial potentials $\phi^{2n}$
\be
 n_s \approx 1- {n+1 \over  N_e} \ .
\ee

This means that the brane inflation spectral index $n_s$ at small $r$ coincides with $n_s$ of inflation in a theory with a polynomial potential $\phi^{2n}$ with
\be\label{ntop}
n \, \Leftrightarrow  \, { 7-p \over  9-p}  \ .
\ee
 The quartic  brane inflation for D3-$\overline{\rm D3}$ model  at small $r$ has the same $n_s$~\cite{Kachru:2003sx}  as the one for $\phi^{4/3}$: 
 \be\label{3}
n_s = 1-{5\over 3 N_{e}}  \qquad  \, \Leftrightarrow  \, \qquad n_s (\phi^{4/3}) \approx 1- {{2\over 3}+1 \over  N_e} \ ,
\ee
in agreement with with the general equation~\rf{gen} for $p=3$. 

  The quadratic  brane inflation model, with D5-$\overline{\rm D5}$ potential,  at small $r$ has the same $n_s$~\cite{GarciaBellido:2001ky} as inflation in a theory with a linear potential $\phi$, 
\be
 n_s \approx 1- {3 \over 2 N_e}  \qquad  \, \Leftrightarrow  \, \qquad n_s (\phi) \approx 1- {{1\over 2}+1 \over  N_e} \ ,
\ee
in agreement with the general equation~\rf{gen} for $p=5$.\footnote{Note that the models with $V\sim \phi^{2n}$, such as  $ \phi^{4/3}$, $\phi$, and $\phi^{2/3}$, can be described by string theory monodromy models \cite{Silverstein:2008sg,McAllister:2008hb,Dong:2010in,McAllister:2014mpa}.  The corresponding models that we describe give the same values of $n_{s}$, for $n$ related to $p$ by the rules \rf{gen}, (\ref{ntop}), but for smaller range of $r$.}

On the left side in  Fig. \ref{Quartic1} we have an $\alpha$-attractor band which starts at $\phi^2$ and moves down in a straight line. Next is the inverse quartic  brane inflation model  which at small $r$ is in the position corresponding to $\phi^{4/3}$, both for BI and KKLTI models,  and continues straight. The inverse quadratic  brane inflation  at small at the right panel in  Fig. \ref{Quartic1} at small $r$ is in a  position corresponding to $\phi$,  both for BI and KKLTI models, and continues straight. These three models pretty much cover all admissible area in the $n_s-r$ plane below $r<10^{-2}$.

\begin{figure}[!h]
%\vspace*{3mm}
\hspace{-3mm}
\begin{center}
\includegraphics[width=9cm]{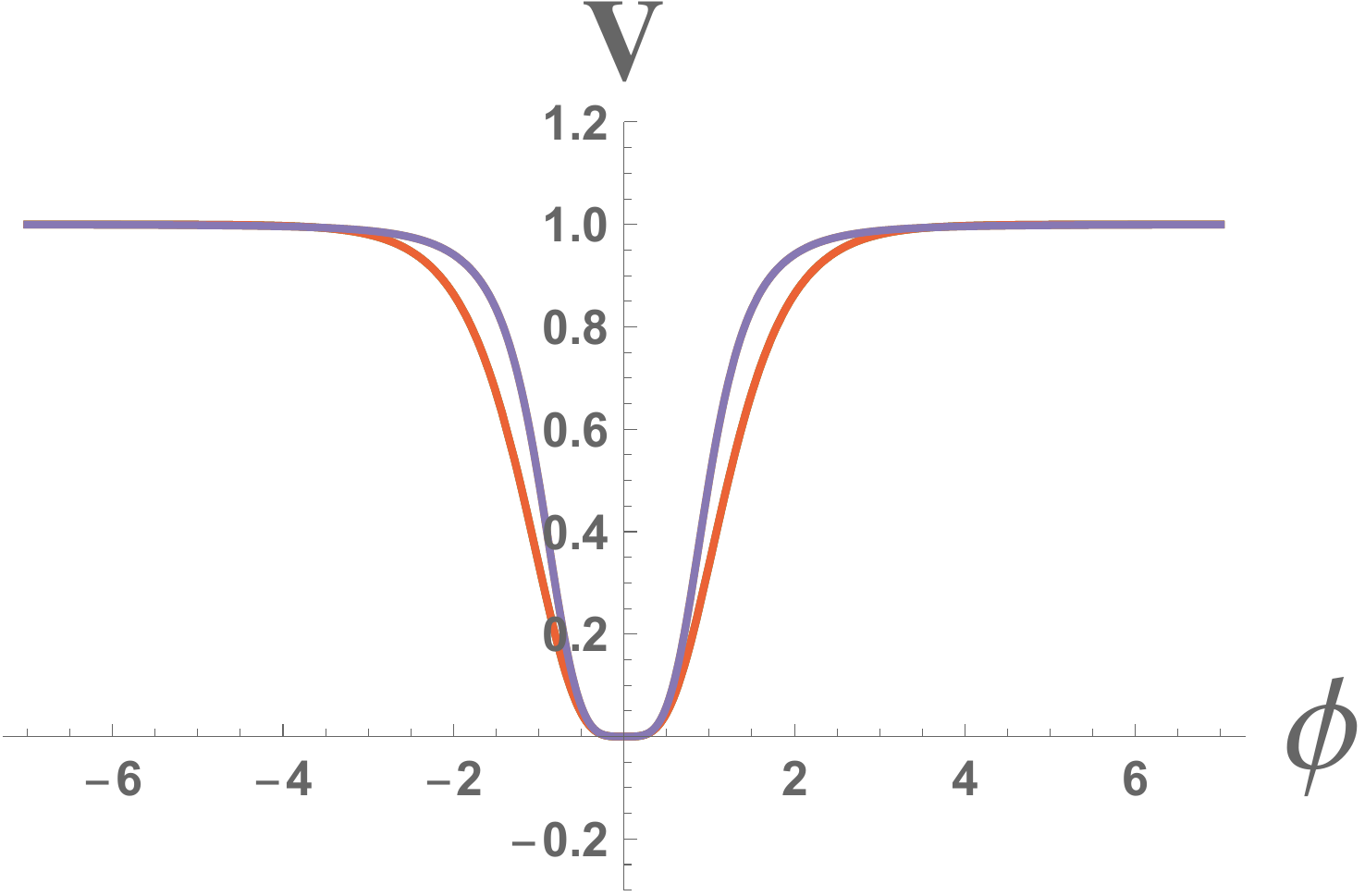}
\caption{\footnotesize  The red line shows the quartic $\alpha$-attractor potential $V= \tanh^4  {\phi \over \sqrt{6 \alpha} } $ with $\alpha =1/6$.  The dark blue line shows  the KKLTI potential $V= {\phi^4\over \phi^4 + m^{4}}$ for $m=1$. In both cases we have plateau potentials at large   $\phi$. For some choices of parameters (e.g. for $\alpha =1/9$ and $m = 1$) these potentials almost coincide, which explains similarity of predictions of these models.}
\label{Compare}
\end{center}
\vspace{0cm}
\end{figure}

To understand the reason of similarity between predictions of $\alpha$-attractors and D-brane inflation, we show in Fig. \ref{Compare} two potentials. One of them is the potential of the $\alpha$-attractor model with $V= \tanh^4  {\phi \over \sqrt{6 \alpha} } $ with $\alpha =1/6$.  The second one is of the inverse quartic D-brane inflation type, $V= {\phi^4\over \phi^4 + m^{4}}$ for $m=1$. This figure shows that in both cases we have plateau potentials at large fields, and for some choice of parameters  (e.g. for $\alpha =1/9$ and $m = 1$) one can even make these potentials look even more similar, even though predictions of these models for $n_{s}$ continue to be slightly different.

The value of $r$ for $\alpha$-attractors depends on $\alpha$,
\be
r\approx {12 \alpha\over N_e^2} \, .
\ee
Meanwhile in brane inflation models the value of $r$ depends on $m$.  In quartic case, using~\cite{Kachru:2003sx} for $m \lesssim 1$ one finds 
\be
  r\approx {4m^{{4\over 3} }\over (3N_e)^{5\over 3}} \ .
\ee
In particular, for $m = 1$, $N_{e} = 50$ we find 
$r\sim  10^{{-3}}$. Using eq. (5.332) of~\cite{Martin:2013tda}, we find a more general expression for $r$ for $V=1-\left(\frac{m}{\phi}\right)^k$,
\be
r\approx 8k^2\left(k(k+2)N_e\right)^{-\frac{2k+2}{k+2}}m^{\frac{2k}{k+2}}.
\label{F}\ee
For an inverse quadratic potential with $k=2$ we find
\be
r\approx   {\sqrt 2\, m \over N_e^{3\over 2}} \ .
\ee
In particular, for $m = 1$, $N_{e} = 50$ we find 
$r   \sim 4 \times 10^{-3}$.

It is instructive to compare these results with the ones presented in Fig.  \ref{Quartic1}. One can see that with the decrease of $r$ the results for $n_{s}$ in the BI and KKLTI models converge to each other at the values of $r$ approximately corresponding to $m \sim  1$, just as one could expect by comparing to each other the potentials of these models. 

Thus here we have presented our  analysis of quartic and quadratic BI  and KKLTI models inside $2\sigma$  region of Planck 2018. As we see, the combination of these models covers the main part of the $2\sigma$ area in the Planck 2018 data in Fig.~\ref{Quartic1}. This is explained by the similarity of potentials of these models illustrated by Fig. \ref{Compare}.  Our results are compatible with the ones found in~\cite {Martin:2013tda}.  

\subsection{\boldmath{Inverse linear case with D6-$\overline{\rm D6}$ potential}} 
In the previous discussion we concentrated on investigation of inverse quadratic and inverse quartic mode, as in the Planck 2018 analysis in \cite{Akrami:2018odb}. Both cases are associated with type IIB string theory, where moduli stabilization was viewed as possible due to KKLT and LVS constructions. However, there was a progress  recently with regard to an uplifting role of the $\overline{\rm D6}$  brane in \cite{Kallosh:2018nrk} and de Sitter vacua in type IIA string theory in \cite{Blaback:2018hdo}.

Therefore we would like  to add here an example of D6-$\overline{\rm D6}$ potential using the general equations above to find out the phenomenology of these models.
This is the case of $k= 7-p = 1$,  
\be
^1 V_{BI}=V_0 \Big [1- \Big ({m\over  |\phi|}\Big ) +\cdots  \Big ]\, , 
\label{V1}\ee
\be
 ^1V_{KKLTI}=V_0 \Big [1+ \Big ({m\over  |\phi|}\Big ) \Big ]^{-1} = V_0 \Big [{ |\phi|\over 
m +  |\phi|} \Big ] \ .
\label{V1exact}\ee
Note that the variable $\phi$ is not a coordinate, but a distance in the moduli space, $\phi = |\phi| \times e^{{i \theta}}$, which is why the potentials depend on $|\phi|$. 

At large $m$, the predictions of the $^1V_{KKLTI}$  model converge  to the predictions of the theory with a simple linear potential $V \sim \phi$. However, at small $m$ it predicts the same value of $n_s$ as the theory with  $V \sim \phi^{2/3}$.

At large $N_e$ and small $r$ (for $m < 1$), the predictions are
\be
 n_s \approx  1- {4 \over  3N_e}  \ , \qquad r\approx  { 8 m^{2\over 3} \over \left(3N_e\right)^{{3\over 4}}}.
\label{lin}
\ee
For $N_{e} = 50$ this gives $n_s \approx 0.973$, which is within the $2\sigma$ Planck area for small $r$.

\section{D-brane Dynamics}

\subsection{String theory analysis}

\begin{figure}[!h]
%\vspace*{3mm}
\hspace{-3mm}
\begin{center}
\includegraphics[width=5cm]{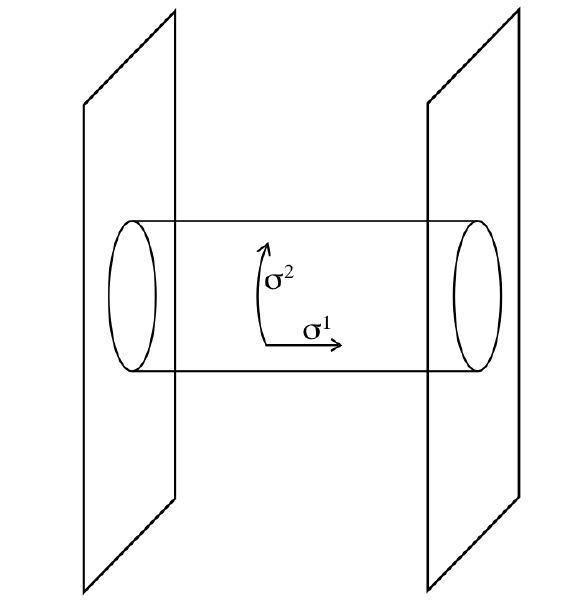}
\caption{\footnotesize  Exchange of a closed string between two D-branes. Equivalently, a vacuum loop of an open string with one end on each D-brane.
}
\label{Joe}
\end{center}
\vspace{0cm}
\end{figure}

To understand the D-brane dynamics in string theory in application to inflation it is useful to consider  the original Polchinski's computation of the energy between two  parallel Dirichlet p-branes at the distance $Y$ from each other~\cite{Polchinski:1996na,Bachas:1998rg}, see Fig.~\ref{Joe}. In eq. (90)  in~\cite{Polchinski:1996na} there is an answer for
two D-branes\footnote{ An open string theory computation of the related diagram is presented in~\cite{Kiritsis:1997hj}.}. Here we present it both for two D-branes, with the  negative contribution from the open string R sector,   as well as for one D-brane-anti-D-brane  with  the  positive contribution from the open string R sector
\bea
A=   \pm  {\cal A}_{\rm R} + {\cal A}_{\rm NS} = 2V_{p+1} \int {dt\over 2t}\, (8\pi^2 \alpha' t)^{-{p+1\over 2}}
e^{- t{Y^2\over 2\pi \alpha'}} \,
\frac{ \mp f_2(q)^{8}
+ f_3(q)^8 - f_4(q)^8 }{2 f_1^8(q)},
\eea
where  $q =e^{-\pi t}$ and all functions $f_i(q)$ are defined in eq. (49) in~\cite{Polchinski:1996na} and $t$ is world sheet modulus. This takes into account that the world-sheet
fermions that are periodic around the cylinder  correspond to R-R
exchange, while the ones which are anti-periodic come from NS-NS exchange.

These three terms for two Dp-branes sum to zero by the `abstruse
identity,' since in this case  the open string spectrum is supersymmetric. But in case of D-brane-anti-D-brane, when R-sector has a different sign,  they sum up since
\bea
{\cal A}_{\rm NS}\ =\ - {\cal A}_{\rm R} .
 \eea 
Thus for  two D-branes
\bea
A_{Dp/ Dp}=    {\cal A}_{\rm R} + {\cal A}_{\rm NS}  &=& 2V_{p+1} \int {dt\over 2t}\, (8\pi^2 \alpha' t)^{-{p+1\over 2}}
e^{- t{Y^2\over 2\pi \alpha'}} 
 \nonumber\\
&& \qquad\qquad\qquad\qquad\qquad \frac{ - f_2(q)^{8}
+ f_3(q)^8 - f_4(q)^8 }{2 f_1^8(q)} =0.
\eea
In terms of the closed string exchange, this
reflects the fact that D-branes are BPS states, the net forces from
NS-NS and R-R exchanges canceling.
For D-brane-anti-D-brane 
\bea
A_{Dp/\bar Dp}=   - {\cal A}_{\rm R} + {\cal A}_{\rm NS} = 2 {\cal A}_{\rm NS} &=& 2V_{p+1} \int {dt\over 2t}\, (8\pi^2 \alpha' t)^{-{p+1\over 2}} e^{- t{Y^2\over 2\pi \alpha'}} \, 
 \frac{  f_2(q)^{8}
 }{ f_1^8(q)}.
\label{full}\eea
We are interested in the limit $t \rightarrow 0$  which is dominated by the lowest lying modes in the closed string spectrum. In such case 
\bea
A_{Dp/\bar Dp}^{t \rightarrow 0}  
 &\sim &  V_{p+1}  \int{dt\over t} (2\pi
t)^{-(p+1)/2} (t/2\pi\alpha')^4 e^{- t{Y^2\over8\pi^2 \alpha'^2}} 
\nonumber\\
\cr
&=& V_{p+1} 4\pi (4\pi^2\alpha')^{3-p} G_{9-p}(Y^2).
\label{smallt}\eea 
Here $G_{9-p}(Y^2)$ is a massless propagator in the Euclidean $9-p$ dimension. The corresponding equation is
\be
\Delta^2_{9-p} G_{9-p}(Y^2)= C \delta ^{9-p} (Y)\, , \qquad G_{9-p}(Y^2) \sim \Big [ c_1 + {c_2\over Y^{7-p}}\Big] ^{-1}.
\ee
For example, for D3 case it is $\Delta^2_{6} G_{6}(Y^2)= C \delta ^{6} (Y)$ and 
\bea
A_{D3/\bar D3}^{t \rightarrow 0} 
 &\sim &  \Big ( c_2 + {c_1\over Y^{4}}\Big)^{-1} \sim b_2 - {b_1\over Y^{4}} +\cdots.
\label{both}\eea 
Thus for two D-branes the net force  between Ramond-Ramond repulsion and gravitational plus dilaton attraction cancels. For D-brane-anti-D-brane the force between Ramond-Ramond attraction and gravitational plus dilaton attraction doubles. 

 At large distances $Y$ one can use an approximated expression for the cylinder amplitude at \rf{smallt}.
 At short distances $b_1 - {b_2\over Y^{4}}$ blows up, but it means that our approximation where only low lying string states are taken into account is not valid and the full tower of string states contributes as shown in eq.~\rf{full}. 

From the perspective of  string theory computation with the result in \rf{full} it would be confusing to use the concept of the brane-anti-brane annihilation. D-branes and anti-D-branes are not particles with the opposite charge  which annihilate and disappear. When the distance between them is small one should not use the approximation $t\rightarrow 0$ which allows to use the harmonic function in eq.~\rf{smallt} and the corresponding potential energy of the brane-anti-brane system. It is the difference between particles and strings which becomes essential at small distances which has to be taken into account in attempts to  provide an interpretation of this physical system.

In application to cosmology we will be interested in both potentials in eq.~\rf{both}. But we will be particularly interested in the region where the difference between these two is not significant.

\subsection{On ${\rm D3}/\overline {\rm D3}$ potential in the space-time picture}

Now we follow the strategy in KKLMMT paper  \cite{Kachru:2003sx} where in Sec. 2 and in Appendix B a computation of the D3/$\overline {\rm D3}$ potential in warped geometries is proposed.  This is also related to a discussion above were we followed \cite{Polchinski:1996na,Bachas:1998rg} in their computation of the cylinder diagram directly in string theory. It is also useful to follow \cite{Lorenz:2007ze} where the same procedure is explained conveniently, not necessarily requiring a warped 5d geometry.  The review  of inflation in string theory in \cite{Baumann:2014nda} is also helpful here. Note that here we assume that we already have a space-time picture. This means that the full string theory computation in \rf{full} is already approximated by the case where $t\rightarrow 0$ limit is taken. But we will see a different way how the harmonic function in the $9-p=6$ Euclidean space shows up and defines the potential of the D3/$\overline {\rm D3}$ system.

D3-brane is perturbing the background and we calculate the resulting energy of the $\overline {\rm D3}$-brane in this perturbed geometry. This  gives the same answer for the potential energy of the brane-anti-brane pair.  We start with 10d geometry
\be
ds^2= h^{-1/2} ds_4^2 + h^{-1/2} ds_4^2\, ,  \qquad F_5 = \partial_ r h^{-1} \ .
\ee
In case that the moving D3-brane is at a position $r_1$ at a radial location in the six-dimensional space and the $\overline {\rm D3}$-brane is at a fixed position $r$ the corresponding harmonic function in the six-dimensional space is given in KKLMMT as
\be
h+\delta h= R^4\Big ({1\over r^4} +{1\over N} {1\over r_1^4}\Big ).
\ee
Here $R$ is a characteristic length scale of the $AdS_5$  geometry, and  $N$ is the five-form charge.
 This expression is valid for strongly warped geometries. In~\cite{Lorenz:2007ze} a  general form of a harmonic function is used, namely for a position of the moving brane $\phi$, which has a canonical kinetic term following from the D-brane action
\be
\tilde h(\phi)= c_2 + {c_1\over \phi^4} \ ,
\ee
where $c_1, c_2$ are some constants. This choice is more in the spirit of the analysis in the previous section where an inverse  harmonic function in a six-dimensional Euclidean space for the D3/$\overline {\rm D3}$ potential is an approximation to the stringy computation
of the cylinder diagram in Fig. \ref{Joe}.  

The Born-Infeld and Chern-Simons actions of the $\overline {\rm D3}$-brane in the background of a moving D3-brane are given by the following expression:
\be
S_{D3/\overline {D3}}=-  \int  \Big [ T(\phi) \sqrt{ 1+ {1\over T(\phi)} g^{\mu\nu} \partial_\mu \phi \partial_\nu \phi} + T(\phi)  \Big ] \sqrt {-g}\, d^4 x.
\ee
Here
\be
T(\phi)= {T_3\over \tilde h(\phi)}.
\ee
Therefore, the action representing D3/$\overline{\rm D3}$ interaction for slow velocities leads to the action of the inflaton field
\be
{\cal L}= -{1\over 2} \dot \phi^2 - V_{inf}( \phi) \ ,
\ee
where
\be
V_{inf}( \phi)=V_0  \Big (1 + {c\over \phi^4}\Big )^{-1}= V_0 \Big ({\phi^4 \over   c+  \phi^4}\Big) \ .
\label{infl}\ee
From the D-brane action $V= {T_3\over h(\phi)}$. Thus
\be
V_0\equiv  {T_3\over c_2}\, , \qquad c\equiv {c_1\over c_2}  \ .
\ee
At $\phi \rightarrow \infty$ the potential $V_{inf }\rightarrow V_0$, at $\phi \rightarrow 0$ the potential $V_{inf }\rightarrow 0$. At $\phi=0$ the only potential which is left is a KKLT potential corresponding to an uplift of dS vacuum, as defined in \rf{KKLT}.

\section{D-brane Inflation in String Theory/de Sitter Supergravity}
We propose to use a shift symmetric \K\, potential of the form
\be
K=-3\log (T+\bar{T}-(\Phi+\bar{\Phi})^2).
\label{ssK}\ee
This kind of shift symmetric \K\, potentials were used in the past for a class of string theory cosmological models in the context of $K3\times {T^2\over \mathbb{Z}_2}$ compactification~\cite{Dasgupta:2002ew,Hsu:2003cy,Hsu:2004hi,Kallosh:2007ig,Haack:2008yb}. See also  review of string cosmology models with unwarped branes in~\cite{Baumann:2014nda}. It is interesting that the relevant stringy D3/D7 models of inflation, called D-term inflation in its supergravity version, generically has $n_s=0.98$ and is now  practically ruled out. However, its investigation gave  us  a useful tool: shift symmetric \K\,potential~\rf{ssK}. 

\subsection{Inflaton shift symmetry and quantum corrections}\label{qcorrection}
It is known that at the classical level in string theory/supergravity one can find situations, like a choice of compactification manifold, when the shift symmetry as shown in eq.~\rf{ssK} is possible. It was actually derived in~\cite{Hsu:2004hi} in case of $K3\times {T^2\over \mathbb{Z}_2}$ compactification, using the special \K\,geometry and the corresponding  holomorphic section $\Omega= \Big (X^\Lambda, F_\Lambda = {\partial F\over \partial X^\Lambda}\Big )$, which, in the special coordinate symplectic frame, is expressed in terms of a prepotential $F$ depending on closed  string moduli $(s,t,u)$ as well as open string moduli, including a position of D3 brane.

 However, quantum corrections may break this symmetry to some extent. The corresponding studies were performed in~\cite{Berg:2004ek,McAllister:2005mq}, mostly in the context of the stringy version of the D3/D7 brane inflation. The situation there may be summarized as follows with regard to detailed studies in D3/D7 brane inflation~\cite{Haack:2008yb} in notation of that paper, where $s= C_4 - i Vol (K3)$, $t$ is the torus complex structure, and $u$ is the axion-dilaton. With account taken of quantum corrections to gauge coupling of D7 brane, the non-perturbative superpotential dependence on the position of D3 brane $y_3$ was given in the form where quadratic and quartic corrections to 
 \be
 W_{np}= A\bigl( 1-\Delta (t_0) y_3^2- \Sigma(t_0) y_3^4 +\cdots\bigr) e^{-ias}\ ,
 \label{np}\ee
 where
$
\Delta (t)= -{2\pi^3 \over 3 c} [ E_2(t) + \vartheta_3^4 (0, t) + \vartheta_4^4 (0, t)]$,
and $\Sigma(t_0)$ is a  function of $\vartheta$ and $E_2$, depending on complex structure modulus $t$, given in eq.~(F.18) of ~\cite{Haack:2008yb}. 
Here $c$ is some group theoretical factor, $\vartheta(\nu, t)$ represent  string theory theta functions. The function $E_2(t) $ was introduced in~\cite{Berg:2005ja}.
In~\cite{Haack:2008yb} the effect of this dependence of the superpotential on the mobile D3 position $y_3$ was taken into account in the potential. As a result, in addition to a standard supergravity D-term potential, there are quantum corrections  of the form of a mass term with a parameter $m^2$ and a quartic term with the parameter $\lambda$.
 Few examples were studied and plotted in  Figs. 4 and 7 in~\cite{Haack:2008yb}. These examples have shown some region of complex structure modulus $t$ where the corrections to $m^2$ as well as to $\lambda$ are small. The conclusion was made that with some fine-tuning of the value of $t$, defined by fluxes, it was possible to make quantum corrections to the potential of D3/D7 brane inflation small. But the problem of  D3/D7 brane inflation with and without quantum corrections is that none of  these models fit the data from Planck 2018.
 
 For D-brane inflation models with $K3\times {T^2\over \mathbb{Z}_2}$ compactification, quantum corrections to superpotential have not been studied. If the calculations of quantum corrections associated with gaugino condensation  in D3/D7 brane inflation  would apply also to D-brane inflation models, one would expect that  fine tuning is necessary to make these corrections small.  It would be interesting to investigate this issue  and describe the situations when such corrections are small, since without quantum corrections D-brane inflation models fit the data from Planck 2018 so well.

In particular, one may study the case where KKLT-type volume stabilization  proceeds via a superpotential generated by Euclidean D3-branes \cite{Witten:1996bn}, not by gaugino condensation. 
The nonperturbative effects in absence of a background flux require that the relevant four-cycle satisfies  a topological condition derived in \cite{Witten:1996bn}. 
However, it was shown in \cite{Gorlich:2004qm,Bergshoeff:2005yp} that these topological conditions are  changed in the presence of flux.  In this case the one-loop correction comes from an instanton fluctuation determinant, which has not been computed in the context of the cosmological models that we study here. It would be important to find out whether such  computation can be performed and what it would entail.

\subsection{A nilpotent multiplet in $\alpha$-attractors} 

An additional tool we use here is a nilpotent multiplet $S$, representing an uplifting $\overline {\rm D3}$-brane,~\cite{Ferrara:2014kva,Kallosh:2014wsa,Bergshoeff:2015jxa,Kallosh:2015nia}. de Sitter supergravity~\cite{Bergshoeff:2015tra,Hasegawa:2015bza} is a local version of non-linearly realized global Volkov-Akulov supersymmetry~\cite{Volkov:1972jx}. 
Using the nilpotent multiplet we will build de~Sitter supergravity for the brane inflation models compatible with the data. An important ingredient of cosmological models in dS supergravity 
is the \K\, metric $K_{S\bar S}= \partial_S \partial_{\bar S} K$ of the nilpotent superfield $S$, which depends on other moduli 
\be
 K_{S\bar S} (\Phi, \bar \Phi) S\bar S.
\label{ssK1}\ee
It has been observed in the past~\cite{McDonough:2016der,Kallosh:2017wnt} that the \K\,metric of the nilpotent superfield might carry the information about the inflationary potential. For example, it was shown in~\cite{Kallosh:2017wnt} that the simplest $\alpha$-attractor model with the potential 
\be 
V(\phi) = \Lambda + m^2 \tanh^2 {\phi\over \sqrt{6\alpha}}
\ee
 can be presented by a particular dS supergravity with the following nilpotent superfield geometry
\be
K_{S\bar S} (Z, \bar Z)  = {W_0^2\over |F_S|^2 + V_{inf} (Z ,\bar Z)}
= {W_0^2\over |F_S|^2 + m^2 Z \bar Z},
\ee
where $F_S=D_SW$. Here $Z$ is a disk variable of the hyperbolic geometry, $m^2 Z \bar Z=m^2\tanh^2 {\phi\over \sqrt{6\alpha}} = V_{inf} (Z ,\bar Z)$ and the cosmological constant at the exit from inflation is given by the difference between two constants, $\Lambda= |F_S|^2- 3 W_0^2>0$.  

In case of Dp-brane inflationary models we will show below that the dependence of the nilpotent field geometry $K_{S\bar S} $ on the inflaton superfield $(\Phi, \bar \Phi)$ has an interesting explanation. It comes  in the context of the KKLMMT construction combined with the recent investigation in~\cite{Kallosh:2018nrk} of the dictionary between string theory models with local sources in ten dimensions and the four-dimensional de Sitter supergravity.

\section{On stringy origin of the nilpotent geometry  $K_{S\bar S} (\Phi, \bar \Phi)$ }

Recently the  dictionary between string theory models and $K$ and $W$ for dS supergravities with closed string moduli was established in~\cite{Kallosh:2018nrk}. In case of open string moduli the analogous analysis was not performed yet. Here we will consider a very particular situation, known from cosmology,  where it is possible to identify the relevant geometry of the nilpotent multiplet from the first principles of string theory with D-branes.

Consider modifications of $K$ and $W$ due to the presence of the nilpotent multiplet, 
\be
K^{\rm new}(z^i, \bar z^i; S, \bar S)= K(z^i, \bar z^i) + K_{S\bar S}(z^i, \bar z^i) S\bar S,
\ee
\be
W^{\rm new} (z^i, S)= W (z^i ) + \mu^2 S\,.
\label{W}\ee
Here the superpotential has a simple dependence on $S$ as in \rf{W}. When $z^i, \bar z^i$  are closed string moduli we have shown in~\cite{Kallosh:2018nrk} why $K_{S\bar S}(z^i, \bar z^i)$ is computable: for each set of ingredients in 10d of the so-called `full-fledged string theory models' one can compute $K_{S\bar S}(z^i, \bar z^i) $ in 4d as a function of the overall volume, the dilaton and the volume moduli of the supersymmetric cycles on which the $\overline {\rm Dp}$-branes are wrapped. 

Since the nilpotent multiplet does not have a scalar component, the new potential has an additional term, but it still depends on the same closed string moduli.
The new F-term potential acquires {\it  an additional nowhere vanishing positive term},  associated with Volkov-Akulov non-linearly realized supersymmetry
\be
V^{\rm new}(z^i, \bar z^i) = V(z^i, \bar z^i) + e^{K(z^i, \bar z^i)} |D_S W|^2\,,   
\ee
where
\be
 |D_S W|^2\equiv  D_S W  K^{S\bar S}(z^i, \bar z^i)\overline{D_{S} W}\,,
\label{newV}\ee
and 
\be
V(z^i, \bar z^i)=e^{K(z^i, \bar z^i)}( |D_i W|^2 -3|W|^2) \ee
 is the standard supergravity potential without the nilpotent multiplet (without the $\overline {\rm D3}$ brane in string theory).  It is important to stress here that there is a dictionary between string theory models in ten dimensions described by supergravity with fluxes and local sources, Dp-branes and Op-planes. Upon compactification on calibrated manifolds these string theory models lead to specific choices of $K$ and $W$ in four-dimensional supergravity, see~\cite{Kallosh:2018nrk} and references therein.

The reason why the nilpotent field metric,  $K_{S\bar S}(z^i, \bar z^i) $ is computable in string theory is that on one hand, the corrected potential due to presence of $\overline {\rm Dp}$-brane has a simple dependence on moduli, $K^{S\bar S}(z^i, \bar z^i) $ under condition that $D_SW|_{S=0}= \mu^2$, as shown in eqs. \rf{W}, \rf{newV}. Thus, the extra potential has a simple dependence on geometry of the nilpotent superfield
\be
V^{\rm new} - V= e^{K(z^i, \bar z^i)}\mu^4 K^{S\bar S}(z^i, \bar z^i).
\label{general}\ee
 On the other hand, the addition to potential due to $\overline {\rm Dp}$-brane action can be inferred from the knowledge of the bosonic $\overline {\rm Dp}$-brane action. By comparing these two we have identified in~\cite{Kallosh:2018nrk} the values of $K_{S\bar S}(z^i, \bar z^i) $ as functions of closed string moduli, 
 \be
V^{\rm new} - V= e^{K(z^i, \bar z^i)}\mu^4 K^{S\bar S}(z^i, \bar z^i) = V_{\overline {Dp}} (z^i, \bar z^i) .
\ee
Here the corresponding action for the $\overline{\rm Dp}$ brane wrapped on a $p-3$ supersymmetric cycle is given by the following expression, and it depends on various closed string moduli, including the volume of the supersymmetric cycles,
\be\label{eq:newterm}
V_{\overline{Dp}} = 2 N_{\overline{Dp},\alpha} T_{Dp} \int_{\Sigma_\alpha} d^{p-3}\xi\, e^{-\varphi} \sqrt{\text{det} ( G+B-2\pi \alpha' F)}\,.
\ee
More details about this action can be found in~\cite{Kallosh:2018nrk}. This leaves us  with the dictionary between the nilpotent field geometry in presence of a pseudo-calibrated $\overline {\rm Dp}$-brane and string theory models with closed string moduli,
\be
K_{S\bar S}(z^i, \bar z^i) = \mu^4 {e^{K(z^i, \bar z^i)}\over V_{\overline {Dp}}(z^i, \bar z^i) }.
\ee
Here we study a particular case of the computation of  $K_{S\bar S}(z^i, \bar z^i; \Phi, \bar \Phi) $, where $\Phi$ is an open string moduli. The new relation between the energy and geometry is an analog of eq.~\rf{general}
\be
V^{\rm new} - V= e^{K(z^i, \bar z^i; \Phi, \bar \Phi )}\mu^4 K^{S\bar S}(z^i, \bar z^i; \Phi, \bar \Phi)\, .
\label{generalOpen}\ee
Thus if we know the dependence of the potential on moduli which is added to the standard supergravity action via a nilpotent field, we can find the geometry using eq.~\rf{generalOpen}.
The corresponding geometry of the nilpotent superfield is determined by the total potential
\be
K_{S\bar S}(z^i, \bar z^i; \Phi, \bar \Phi) = \mu^4\, {e^{K(z^i, \bar z^i ; \Phi, \bar \Phi)}\over V^{\rm new} - V}\ .
\label{Genmetric}\ee

\subsection{KKLT uplift}

A manifestly supersymmetric version of the KKLT uplifting was proposed in the form in which the $\overline {\rm D3}$-brane is represented by a nilpotent multiplet $S$ with $S^2=0$, corresponding to Volkov-Akulov  non-linearly realized supersymmetry~\cite{Ferrara:2014kva,Kallosh:2014wsa,Bergshoeff:2015jxa,Kallosh:2015nia}. In this case the new $K$ and $W$ are given by (in unwarped case)
\be
\begin{aligned}\label{KKLT}
&K= -3\log\left(T+\bar T\right)+  S \bar S \, , \\ 
 &W= W_0 + A \exp(-a T) + \mu^2S\,, \end{aligned}
\ee
and \be
 V_{up}=V^{\rm new} - V= \left. e^K |D_S W|^2 \right|_{S=\bar S=0}= {\mu^4\over   \, (T+\bar T )^3}\,.
\label{up}\ee
This is in agreement with using only closed string moduli and $V_{\overline {D3}}$ action. At present there is a consensus 
that eqs.~\rf{KKLT} and~\rf{up} represent a manifestly supersymmetric version of the KKLT uplift. It involves a nilpotent multiplet representing an $\overline {\rm D3}$ brane in the framework of de Sitter supergravity  with a non-linearly realized supersymmetry.

\subsection{Inflationary uplift}

Here we consider the situation where at the end of inflation the uplifting energy is due to an $\overline {\rm D3}$ brane which is at some fixed point in the manifold~\cite{Kachru:2003sx,Kachru:2003aw},  for 
example  on a top of an O3-plane, as discussed more recently in~\cite{Kallosh:2015nia}. 
 
Our new proposal here is to look for a combination of potentials due to KKLT uplift to dS vacua, and an additional uplift by the inflationary energy depending on open string modulus.
In case of D3-brane inflation, the new term depends on the energy of D3/$\overline {\rm D3}$ interaction. Our proposal means that the corresponding geometry of the nilpotent superfield will be defined by the total potential
\be
V^{\rm new} - V=  V_{\overline D_3} (z^i, \bar z^i)+ V_{D3/\overline {D3}} (z^i, \bar z^i; \Phi, \bar \Phi).
\label{Energy}\ee
In addition to $V_{\overline{D3}}$   we have now  added the energy of the   $V_{D3/\overline {D3}}$ system depending on open string modulus. In such case, 
it follows from~\rf{Genmetric} that
\be
K_{S\bar S}(z^i, \bar z^i; \Phi, \bar \Phi) = \mu^4 {e^{K(z^i, \bar z^i ;\Phi, \bar \Phi)}\over V_{\overline D_3} + V_{D3/\overline {D3}}(\Phi, \bar \Phi)}.
\label{metric}\ee
This is our definition of the  inflationary uplift in the case of D3/$\overline {\rm D3}$ inflation.  We will demonstrate below that it is very useful for inflationary model building.

\section{ Brane Inflation with volume stabilization in dS supergravity}

\subsection{D-brane inflaton potentials}

 Our proposal for a  supersymmetric version of D-brane inflationary models is based on shift-symmetric \K\,potential~\rf{ssK} and on an inflationary potential of the type~\rf{infl}
\be
V_{inf} (\Phi, \bar \Phi) = V_0 \Big ({[-i (\Phi -\bar \Phi )]^{7-p}\over c+[-i (\Phi -\bar \Phi )]^{7-p}}\Big).\label{vinf}
\ee
It could also be a potential of the form
\be
V_{inf} (\Phi, \bar \Phi) = V_0 \Big ( 1- { b\over [-i (\Phi -\bar \Phi) ]^{7-p}} +\cdots\Big) \ ,
\ee
where the terms with $\cdots$ have to be added to remove the singularity at $\Phi -\bar \Phi$. We will see below that the cosmological evolution with strong volume modulus stabilization as proposed in~\cite{Kallosh:2011qk,Linde:2011ja,Dudas:2012wi} works well for the inflationary models we study below. Namely we can use either KKLT type volume stabilization assuming that $m_{3/2} >H$ to avoid volume destabilization during inflation~\cite{Kallosh:2004yh}, or using the KL mechanism with two exponents~\cite{Kallosh:2004yh,BlancoPillado:2005fn,Kallosh:2014oja}. In both cases the process of inflation does not affect the volume modulus stabilization and vice versa, inflation is not affected by the volume modulus stabilization.  One should note that the geometric approach used in our paper may impose certain constraints on the gravitino mass, which should be taken into account in the model building~\cite{Hasegawa:2017nks}.

\subsection{Unifying inflation and strong volume stabilization}
We consider a general theory of volume stabilization in combination of the inflationary potential and discuss the back-reaction of the inflaton potential on the moduli. We use the following set of K\"ahler and superpotential 
\begin{align}
K=&-3\log (T+\bar{T}-(\Phi+\bar{\Phi})^2)+\frac{S\bar{S}}{(T+\bar{T}-(\Phi+\bar{\Phi})^2)^\beta (1+f(\Phi,\bar{\Phi}))},\label{KKKLT}\\
W=&W(T)+\mu^2 S,\label{WKKLT}
\end{align}
where $T=\rho+i\sigma$ is a volume modulus multiplet, $\Phi=\chi+i\phi$ is an inflaton multiplet, $S$ is an $\overline {\rm D3}$ nilpotent multiplet, respectively. $\beta=0,1$ depends on where the $\overline{\rm D3}$ is, and in the warped case, $\beta=1$, in the unwarped case, $\beta=0$. The K\"ahler coupling $f(\Phi,\bar{\Phi})$ gives rise to inflaton potential, see~\rf{metric}.
The scalar potential is given by
\be
V_{\rm inf}= V(T,\chi) +  {\mu^4 f(\Phi,\bar{\Phi})\over   (2\rho-4\chi^2)^{3 - \beta}}\, .
\ee
In all models to be considered, inflation occurs along a stable inflationary trajectory $\sigma= \chi = 0$. We will also consider $f(\Phi,\bar{\Phi}) = F(- i(\Phi-\bar{\Phi})/2)$. In that case, the general expression for the potential of the inflaton field $\phi$ and of the volume modulus $\rho$ is given by
\be
V_{\rm inf}(\rho,\phi)= V(\rho) +  {\mu^4F(\phi)\over   (2\rho)^{3 - \beta}}\, .
\ee
In what follows, we will consider two models where the potential $V(\rho)$ ensures strong stabilization of the volume modulus $\rho$ near its post-inflationary value $\rho_{0}$, such that during inflation one has $\rho \approx \rho_{0}$. Also, one can ensure that the post-inflationary vacuum energy $V(\rho_{0}) = \Lambda \sim 10^{{-120}}$ is many orders of magnitude smaller than the inflaton potential. In that case, the potential during inflation can be represented by a very simple expression
\be\label{generalinf}
V_{\rm inf}= {\mu^4F(\phi)\over   (2\rho_{0})^{3 - \beta}}\, .
\ee
This expression shows that one can easily combine strong moduli stabilization with construction of inflationary models with arbitrary potentials \rf{generalinf}. Similar, but slightly more complicated methods were  used in the past in the phenomenological inflationary models  without nilpotent superfield  \cite{Kallosh:2011qk,Linde:2011ja,Dudas:2012wi}.

\subsection{D-brane inflation and strongly stabilized KKLT model}
We consider the KKLT moduli stabilization with inflationary potential and discuss the back-reaction of the inflaton potential on the moduli. We use the following set of K\"ahler and superpotential \footnote{We do not assume that $A$ in the superpotential has a significant dependence on $\Phi$.} :
\begin{align}
K=&-3\log (T+\bar{T}-(\Phi+\bar{\Phi})^2)+\frac{S\bar{S}}{(T+\bar{T}-(\Phi+\bar{\Phi})^2)^\beta (1+f(\Phi, \bar{\Phi}))},\label{KKKLT2}\\
W=&W_0-Ae^{-aT}+\mu^2 S,\label{WKKLT2}
\end{align}
where $T=\rho+i\sigma$ being a volume modulus multiplet, $\Phi=\chi+i\phi$ being an inflaton multiplet, $S$ being an $\overline {\rm D3}$ nilpotent multiplet, respectively. $\beta=0,1$ depends on where the $\overline{\rm D3}$ is, and in the warped case, $\beta=1$, in the unwarped case, $\beta=0$. The K\"ahler coupling $f(\Phi,\bar\Phi)$ gives rise to inflaton potential, see~\rf{generalinf}.

\begin{figure}[!h]
%\vspace*{3mm}
\hspace{-3mm}
\begin{center}
\includegraphics[width=9cm]{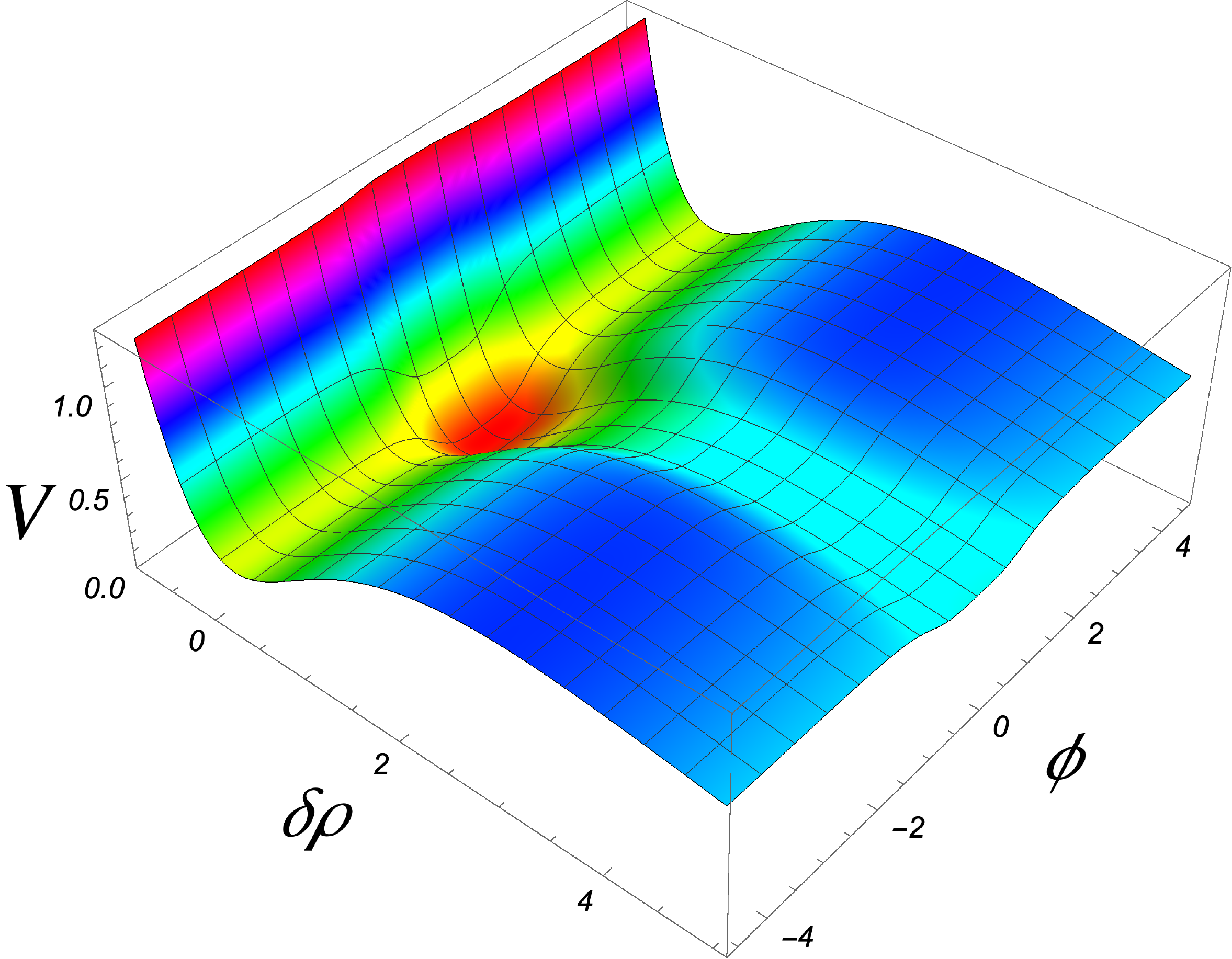}
\caption{\footnotesize  The potential of the volume modulus $\delta\rho$ and the inflaton $\phi$ in KKLT case, \eqref{KKKLT2},\eqref{WKKLT2}. $V_{\rm inf}$ is chosen to be the form $V_{\rm inf}=V_0(1+\frac{\mu^4}{\phi^4})^{-1}$. Inflation is realized along the straight valley of the $\phi$ direction, while the volume $\delta\rho$ is stabilized.  In realistic models one should take $V_{0} \sim 10^{{-10}}$; here we took $V_{0}=1$ just for illustration. The shape of the potential along the inflationary valley in the $\phi$ direction  is shown by the dark blue line in Fig. \ref{Compare}.}
\label{KKLTinfl}
\end{center}
\vspace{0cm}
\end{figure}

Let us first consider the simplest case $f=0$ and $\chi=0$. In this case, the scalar potential is simply given by
\begin{equation}
V_{KKLT}=\frac{\mu^4}{(2\rho)^{3-\beta}}+\frac{aAe^{-a\rho}(-3W_0+Ae^{-a\rho}(3+a\rho))}{6\rho^2},
\end{equation}
where we have minimized with respect to $\sigma$ and $\rho_0$ is the value at which $D_TW=0$ is satisfied. We denote the minimum of this potential as $\rho=\rho_0$, which satisfies
\begin{equation}
W_0=\frac{1}{3}Ae^{-a\rho_0}(3+2a\rho_0)-\frac{2^{-2+\beta}e^{a\rho_0}(-3+\beta)\mu^4\rho_0^{\beta-1}}{aA(2+a\rho_0)}.
\end{equation}
We assume that $f(\chi)=0$ at the minimum of $\chi$. From (almost) vanishing cosmological constant condition, we find
\begin{equation}
\mu^4\simeq\frac{2^{2-\beta}a^2A^2e^{-2a\rho_0}\rho_0^{2-\beta}(2+a\rho_0)}{3(-1+\beta+a\rho_0)}.
\end{equation} 
For simplicity, let us choose $\beta=1$, and one finds 
\begin{equation}
m_{3/2}=\frac{2Ae^{-a\rho_0}(1+a\rho_0)}{6\sqrt{2}\rho_0^{3/2}}.
\end{equation}

Tuning on the inflaton dependence $f$, the potential minimum of $\rho$ is no longer $\rho=\rho_0$. Let us call the deviation from $\rho_0$ as $\delta\rho$. We expand the potential with respect to $\delta\rho$ and minimize it, which gives the following effective potential,
\begin{align}
V_{\rm eff}=&\frac{24a m_{3/2}^2\rho_0^4(2+a\rho_0)f}{(1+a\rho_0))^2}-\frac{48m_{3/2}^2\rho_0^2(2+a\rho_0)^2f^2}{a(1+a\rho_0)^2(3+2a\rho_0)}\nonumber\\
=&V_{\rm inf}-\frac{2V_{\rm inf}^2(1+a\rho_0)^2}{3a^3\rho_0^3(3+2a\rho_0)m_{3/2}^2}.
\end{align}
Here we have set $\chi=0$, which is justified below, and defined $$V_{\rm inf}=\frac{24a m_{3/2}^2\rho_0^4(2+a\rho_0)f}{(1+a\rho_0)^2}.$$ The second term is regarded as the deformation of potential from the back-reaction of heavy modulus $\rho$. We find that the back-reaction term has the factor $V_{\rm inf}/{m_{3/2}^2}\sim H^2_{\rm inf}/m_{3/2}^2$. We remind here that KKLT uplift is consistent with inflation only if the height of the barrier is higher than the scale of inflation, $m_{3/2} > H_{\rm inf}$~\cite{Kallosh:2004yh}. Thus, in our case $H^2_{\rm inf}/m_{3/2}^2$ must be small (i.e. SUSY breaking must be higher than inflation scale), which also protects the system from back-reaction.

Finally, let us comment on the stability of sinflaton $\chi$. Near the minimum $\delta\rho\sim0$, the mass of sinflaton is given by
\begin{equation}
m_{\chi}^2=K^{\Phi\bar\Phi}V_{\phi\phi}=\frac{4(3a^2\rho_0^2m_{3/2}^2+2V_{\rm inf}(1+a\rho_0)^2)}{3(1+a\rho_0)^2}.
\end{equation}
Thus, the sinflaton mass is always greater than inflation scale, and the sinflaton can be set at its origin.

\subsection{KL model}
Next, we consider inflation coupled to KL moduli stabilization \cite{Kallosh:2004yh}. The system is given by
\begin{align}
K=&-3\log (T+\bar{T}-(\Phi+\bar{\Phi})^2)+\frac{S\bar{S}}{(T+\bar{T}-(\Phi+\bar{\Phi})^2)^\beta (1+f(\Phi, \bar{\Phi}))},\label{KKL}\\
W=&W_0-Ae^{-aT}+Be^{-bT}+\delta w+\mu^2 S.\label{WKL}
\end{align}
We focus on the vacuum where 
$$Ae^{-aT}\sim Be^{-bT}\sim W_0\gg \delta w.$$ In particular, we assume the relation $$W_0=A\left(\frac{aA}{bB}\right)^{-\frac{a}{a-b}}\left(1-\frac ab\right).$$ This relation leads to $W=0$ when $\delta w=0=W_T$ is satisfied. One can check that $\sigma=0$ is the minimum.

As previous case, we expand the potential in $\delta\rho=\rho-\rho_0$ up to the quadratic order, where $\rho_0$ is the value of $\rho$ for the case with $\chi=f=0$. By minimizing the potential with respect to $\delta\rho$, we find the following effective potential,
\begin{equation}
V_{\rm eff}=3m_{3/2}^2(1+\cdots)f-\frac{6(3-\beta)^2m_{3/2}^4f^2}{M^2\rho_0}+\cdots,
\end{equation}
where $M=\frac 2 3 aA(a-b)(\frac{aA}{bB})^{-\frac{a}{a-b}}$ is the  mass of $\rho$ at $\delta w=\mu=0$. The ellipses denote the higher order terms suppressed by the factor $m_{3/2}/M$. For $M\gg m_{3/2}$, the leading part of the effective potential is $V_{\rm inf}=3m_{3/2}^2f$. We rewrite the effective potential as
\begin{equation}
V_{\rm eff}\sim V_{\rm inf}-\frac{2 V_{\rm inf}^2}{3M^2}.
\end{equation}
The moduli stabilization during inflation requires $H_{\rm inf} \ll M$, and this condition means that the second term in the effective potential is much smaller than the leading term. Thus, we can safely ignore the correction coming from the back-reaction of the heavy modulus~$\rho$. Note that in KL model the mass $M$ is not related to the mass of gravitino. This allows much greater flexibility with respect to the strength of SUSY breaking.

Finally, we note that the mass of $\chi$ near $\delta\rho=0$ is given by
\begin{equation}
m_{\chi}^2=(8+12f)m^2_{3/2}=8m_{3/2}^2+4V_{\rm inf}.
\end{equation}
Therefore, $\chi$ is stabilized at its origin during inflation with a sufficiently large mass. The shape of the potential in this model is very similar to the one shown in Fig.~\ref{KKLTinfl}.

\section{Discussion}

During the last 15 years the accuracy of determination of the spectral parameter $n_{s}$ increased dramatically. In 2003, after the first WMAP data release,  the combination of all available data suggested that $n_{s} = 0.93 \pm 0.03$~\cite{Spergel:2003cb}. 9 years later, in the 9 year WMAP data release, the result was $n_{s} = 0.972 \pm  0.013$~\cite{Hinshaw:2012aka}.
In the Planck 2013 data release the corresponding number was $n_{s} =  0.9603 \pm 0.0073$~\cite{Ade:2013zuv}. Finally, the Planck 2018 result is  $n_{s} = 0.965 \pm  0.004$~\cite{Aghanim:2018eyx}.

There are many models where by tuning two or three parameters one can get $n_{s} = 0.965 \pm  0.004$, but typically this tuning additionally depends on $r$. As a result, most inflationary models have $n_s$ and $r$ all over the place in the $n_s$~-~$r$ plane. 

One of the rare exceptions is the class of $\alpha$-attractors~\cite{Kallosh:2013yoa}, which  match the  Planck 2018 data without any fine-tuning. Predictions of these models are shown, in the simplest case, in Fig. \ref{Planck}  by  two yellow lines for $N_e=50, 60$. For more general $\alpha$-attractor models,  the predictions for $n_s$ and $r$  are shown in Fig.~\ref{alpha}.  In the large $N_e$ limit, they are given by
\be
^\alpha n_s \approx 1-{2\over N_e} \ , \qquad ^\alpha r \approx {12\alpha \over N_e^{2}} \ .
\ee
These models for small values of $r$ tend to cover the left side of the $1\sigma$ area in the Planck 2018 data, see Figs. \ref{Planck}, \ref{Quartic}, \ref{alpha}, \ref{Quartic1}.

In this paper we revisited D-brane inflation models in string theory in the context of volume stabilization~\cite{Kachru:2003sx,Dvali:2001fw,Burgess:2001fx,GarciaBellido:2001ky}.  At a phenomenological level,  the inflationary models with
\be
^4 V_{\rm inf} =V_0 \Big [ 1+\Big ({m\over \phi}\Big )^{4} \Big]^{-1} \, , \qquad ^2 V_{\rm inf} =V_0 \Big [ 1+\Big ({m\over \phi}\Big )^{2} \Big]^{-1}\, , \qquad ^1 V_{\rm inf} =V_0 \Big [ 1+\Big ({m\over |\phi|}\Big ) \Big]^{-1}
\ee
have many nice  features similar to those of the $\alpha$-attractors. Namely, as shown in \cite{Burgess:2001fx,GarciaBellido:2001ky} and in \cite{Kachru:2003sx}, in inverse quartic case and in inverse quadratic case  they have  universal predictions for $m \lesssim 1$ at large  $N_e$ and small $r$:
\be
^4 n_s \approx  1- {5 \over  3 N_e}   \, , \qquad ^4 r\approx {4m^{{4\over 3} }\over (3N_e)^{5\over 3}}\ ,  \ee
and 
\be
^2 n_s \approx  1- {3 \over  2 N_e}\ ,  \qquad ^2 r\approx {12 \alpha\over N_e^2}   \ .
\ee
We have also added the case of an inverse linear potential
\be
^1 n_s \approx  1- {4 \over  3 N_e}\ ,   \qquad ^1 r\approx  { 8 m^{2\over 3} \over \left(3N_e\right)^{{3\over 4}}}
  \ .
\ee

These models have specific values of $n_s$ which nicely agree with the data, for any choice of parameters. In fact, this property is the same as in $\alpha$-attractor models.

Since $2> {5\over 3} > {3 \over  2}$, we see that the $^4 V_{\rm inf}$ slice of the $n_s$~-~$r$ plane is to the right of the one for the $\alpha$-attractor models, and the predictions for $^2 V_{\rm inf} $ are even more to the right. Adding the linear case, and since $2> {5\over 3} > {3 \over  2}> {4 \over  3}$ we note that this model is even more to the right, compared to $\alpha$-attractors, for the same values of $N_e$.

A combination of $\alpha$-attractors and the D-brane inflation with the inverse quartic potentials for KKLTI and BI models  completely covers the $1\sigma$ sweet spot of the Planck data, and a combination of these models with KKLTI and BI models with quartic potentials almost completely covers the broader $2\sigma$ area of the Planck 2018 data, see Fig. \ref{Quartic1}. The missing in Fig. \ref{Quartic1} linear case is a bit to the right from the quadratic one.

In view of phenomenological importance of the D-brane inflation models, we revisited their stringy origin.  Based on  the original Polchinski's computation of D-brane-D-brane (vanishing) potential as well as the studies of this in~\cite{Kachru:2003sx}, we find that D-brane-anti-D-brane potential leads to a specific dependence on the open string modulus of the geometry of the nilpotent multiplet $K_{S\bar S}(z^i, \bar z^i; \Phi, \bar \Phi) = \mu^4 {e^{K(z^i, \bar z^i; \Phi, \bar \Phi)}\over V_{\overline D_3} + V_{D3/\overline {D3}}(\Phi, \bar \Phi)}$. This is our generalization, to the case of the open string moduli,  of the dictionary between string theory models with local sources in ten dimensions and the four-dimensional de Sitter supergravity studied for  closed string theory moduli in~\cite{Kallosh:2018nrk}.

We  combined this geometric information with the shift symmetric \K\, potential $K=-3\log (T+\bar{T}-(\Phi+\bar{\Phi})^2)$ of the kind known for $K3 \times {T^2\over \mathbb{Z}_2}$ compactification. The resulting de Sitter type supergravity models, with either KKLT or KL volume stabilization, were studied in the regime of  strong volume stabilization. The combined potential of the inflaton and the volume modulus is shown in Fig.~\ref{KKLTinfl}.  One can see a nearly flat inflationary $\chi$ direction, with the D-brane inflaton potential remaining practically unchanged by the presence of the strongly stabilized volume modulus.

To complete this construction, it would be important to study quantum corrections of the type discussed in fiber inflation in~\cite{Cicoli:2008gp,Kallosh:2017wku} and in D3/D7 model in~\cite{Haack:2008yb}, reviewed here in Sec.~\ref{qcorrection}. It will also be important to study quantum corrections in case that KKLT-type volume stabilization  proceeds via a superpotential generated by Euclidean D3-branes \cite{Witten:1996bn,Gorlich:2004qm,Bergshoeff:2005yp}, not by gaugino condensation.
This is a very challenging task, but the phenomenological success of the simple D-brane inflation models considered in this paper suggests that these theories deserve a detailed investigation.  

\
  
 \noindent{\bf {Acknowledgments:}} We are grateful to D. Baumann, S. Ferrara, S. Kachru, J. Maldacena, L. McAllister, E. Silverstein, S. Trivedi, F. Zwirner, 
 and T. Wrase   for stimulating discussions and collaboration on related work.  This work is supported by SITP and by the US National Science Foundation grant PHY-1720397 and by the Simons foundation grant.  
 
 % \newpage

 \appendix
 \section{Other phenomenological models}
We show different phenomenological models having the potential~\eqref{vinf}. The relation between string theory and the following models are not clear. Nevertheless, these phenomenological models are consistent with the current cosmological data and interesting independently of string theory.

\subsection{From flattening mechanism}
First, let us consider the following simple (but non-supersymmetric) system with two scalars $(\chi, \phi)$ with potential
\begin{equation}
V=\frac12m^{2-2p}\phi^{2p}\chi^2+\frac12M^2(\chi-\chi_0)^2 \ ,
\end{equation}
where $M, \ m$ and $\chi_0$ are constants. First, we integrate out the heavy moduli $\chi$, and find the following effective potential
\begin{equation}
V_{\rm eff}=\frac{1}{2}M^2\chi_0^2\left(1+\frac{\mu^{2p}}{\phi^{2p}}\right)^{-1} \ .
\end{equation}
Here we have neglected higher derivative corrections since they are smaller compared to the corrections to potential~\cite{Dong:2010in}.

\subsection{Supergravity model}
Next we consider a supergravity model with
\begin{align}
K=&\frac12 (\Phi+\bar\Phi)^2+S\bar{S}+\left (1+\frac{(\Phi\bar{\Phi})^{p}}{M^{2p}}\right)Y\bar{Y},\nonumber\\
W=&\mu^2S+W_0+\lambda m^{2-p} \Phi^{p}Y,
\end{align}
where $X$ is a nilpotent superfield and $Y$ is a constrained superfield satisfying $XY=0$. Both $X$ and $Y$ do not have dynamical scalars, and $F^X\neq0$ for consistency.\footnote{One may use an unconstrained chiral superfield instead of a constrained one $Y$. Here we use the constrained one for simplicity.} $\Phi=\chi+i\phi$ and $\chi$ is the inflaton. The scalar potential becomes
\begin{align}
V=&(\mu^4-3W_0^2)+\lambda^2m^{4-2p}\frac{|\Phi|^{2p}}{1+\frac{|\Phi|^{{2p}}}{M^{2p}}}\nonumber\\
=&\Lambda^4+\lambda^2(M^p m^{2-p})^2\left(1+\frac{M^{2p}}{|\Phi|^{2p}}\right)^{-1},
\end{align} 
where $\Lambda^4=\mu^4-3W_0^2$ is a cosmological constant which is fine-tuned to be $\mathcal O(10^{-120})$.
We expand the potential with respect to $\chi$, and find
\begin{equation}
V=V_{\rm inf}+\left(4W_0^2+V_{\rm inf}\left(2+\frac{pM^{2p}}{\phi^2(M^{2p}+\phi^{2p})}\right)\right)\chi^2+\mathcal O(\chi^4),
\end{equation}
where 
\begin{equation}
V_{\rm inf}=\lambda^2(m^{2-p}M^p)^2 \left(1+\frac{M^{2p}}{\phi^{2p}}\right)^{-1}.
\end{equation}
Thus, the sinflaton $\chi$ is stabilized at $\chi=0$ during inflation, and we find a single-field inflation with the KKLTI potential.

\bibliographystyle{JHEP}
\bibliography{lindekalloshrefs}

\providecommand{\href}[2]{#2}\begingroup\raggedright\begin{thebibliography}{10}

\bibitem{Akrami:2018odb}
{\scshape Planck} collaboration, Y.~Akrami et~al., \emph{{Planck 2018 results.
  X. Constraints on inflation}},
  \href{https://arxiv.org/abs/1807.06211}{{\ttfamily 1807.06211}}.

\bibitem{Kallosh:2013yoa}
R.~Kallosh, A.~Linde and D.~Roest, \emph{{Superconformal Inflationary
  $\alpha$-Attractors}},
  \href{https://doi.org/10.1007/JHEP11(2013)198}{\emph{JHEP} {\bfseries 11}
  (2013) 198}, [\href{https://arxiv.org/abs/1311.0472}{{\ttfamily 1311.0472}}].

\bibitem{Starobinsky:1980te}
A.~A. Starobinsky, \emph{{A New Type of Isotropic Cosmological Models Without
  Singularity}},
  \href{https://doi.org/10.1016/0370-2693(80)90670-X}{\emph{Phys. Lett.}
  {\bfseries 91B} (1980) 99--102}.

\bibitem{Goncharov:1985yu}
A.~S. Goncharov and A.~D. Linde, \emph{{CHAOTIC INFLATION OF THE UNIVERSE IN
  SUPERGRAVITY}}, {\emph{Sov. Phys. JETP} {\bfseries 59} (1984) 930--933}.

\bibitem{Salopek:1988qh}
D.~S. Salopek, J.~R. Bond and J.~M. Bardeen, \emph{{Designing Density
  Fluctuation Spectra in Inflation}},
  \href{https://doi.org/10.1103/PhysRevD.40.1753}{\emph{Phys. Rev.} {\bfseries
  D40} (1989) 1753}.

\bibitem{Bezrukov:2007ep}
F.~L. Bezrukov and M.~Shaposhnikov, \emph{{The Standard Model Higgs boson as
  the inflaton}},
  \href{https://doi.org/10.1016/j.physletb.2007.11.072}{\emph{Phys. Lett.}
  {\bfseries B659} (2008) 703--706},
  [\href{https://arxiv.org/abs/0710.3755}{{\ttfamily 0710.3755}}].

\bibitem{Carrasco:2015pla}
J.~J.~M. Carrasco, R.~Kallosh and A.~Linde, \emph{{$\alpha$-Attractors: Planck,
  LHC and Dark Energy}},
  \href{https://doi.org/10.1007/JHEP10(2015)147}{\emph{JHEP} {\bfseries 10}
  (2015) 147}, [\href{https://arxiv.org/abs/1506.01708}{{\ttfamily
  1506.01708}}].

\bibitem{Kallosh:2013hoa}
R.~Kallosh and A.~Linde, \emph{{Universality Class in Conformal Inflation}},
  \href{https://doi.org/10.1088/1475-7516/2013/07/002}{\emph{JCAP} {\bfseries
  1307} (2013) 002}, [\href{https://arxiv.org/abs/1306.5220}{{\ttfamily
  1306.5220}}].

\bibitem{Cicoli:2008gp}
M.~Cicoli, C.~P. Burgess and F.~Quevedo, \emph{{Fibre Inflation: Observable
  Gravity Waves from IIB String Compactifications}},
  \href{https://doi.org/10.1088/1475-7516/2009/03/013}{\emph{JCAP} {\bfseries
  0903} (2009) 013}, [\href{https://arxiv.org/abs/0808.0691}{{\ttfamily
  0808.0691}}].

\bibitem{Kallosh:2017wku}
R.~Kallosh, A.~Linde, D.~Roest, A.~Westphal and Y.~Yamada, \emph{{Fibre
  Inflation and $\alpha$-attractors}},
  \href{https://doi.org/10.1007/JHEP02(2018)117}{\emph{JHEP} {\bfseries 02}
  (2018) 117}, [\href{https://arxiv.org/abs/1707.05830}{{\ttfamily
  1707.05830}}].

\bibitem{Ferrara:2016fwe}
S.~Ferrara and R.~Kallosh, \emph{{Seven-Disk Manifold, $\alpha$-attractors and
  B-modes}},  \href{https://arxiv.org/abs/1610.04163}{{\ttfamily 1610.04163}}.

\bibitem{Kallosh:2017ced}
R.~Kallosh, A.~Linde, T.~Wrase and Y.~Yamada, \emph{{Maximal Supersymmetry and
  B-Mode Targets}}, \href{https://doi.org/10.1007/JHEP04(2017)144}{\emph{JHEP}
  {\bfseries 04} (2017) 144},
  [\href{https://arxiv.org/abs/1704.04829}{{\ttfamily 1704.04829}}].

\bibitem{Kachru:2003sx}
S.~Kachru, R.~Kallosh, A.~D. Linde, J.~M. Maldacena, L.~P. McAllister and S.~P.
  Trivedi, \emph{{Towards inflation in string theory}},
  \href{https://doi.org/10.1088/1475-7516/2003/10/013}{\emph{JCAP} {\bfseries
  0310} (2003) 013}, [\href{https://arxiv.org/abs/hep-th/0308055}{{\ttfamily
  hep-th/0308055}}].

\bibitem{Lorenz:2007ze}
L.~Lorenz, J.~Martin and C.~Ringeval, \emph{{Brane inflation and the WMAP data:
  A Bayesian analysis}},
  \href{https://doi.org/10.1088/1475-7516/2008/04/001}{\emph{JCAP} {\bfseries
  0804} (2008) 001}, [\href{https://arxiv.org/abs/0709.3758}{{\ttfamily
  0709.3758}}].

\bibitem{Martin:2013tda}
J.~Martin, C.~Ringeval and V.~Vennin, \emph{{Encyclop\ae dia Inflationaris}},
  \href{https://doi.org/10.1016/j.dark.2014.01.003}{\emph{Phys. Dark Univ.}
  {\bfseries 5-6} (2014) 75--235},
  [\href{https://arxiv.org/abs/1303.3787}{{\ttfamily 1303.3787}}].

\bibitem{Dvali:2001fw}
G.~R. Dvali, Q.~Shafi and S.~Solganik, \emph{{D-brane inflation}},  in
  \emph{{4th European Meeting From the Planck Scale to the Electroweak Scale
  (Planck 2001) La Londe les Maures, Toulon, France, May 11-16, 2001}}, 2001,
  \href{https://arxiv.org/abs/hep-th/0105203}{{\ttfamily hep-th/0105203}}.

\bibitem{Burgess:2001fx}
C.~P. Burgess, M.~Majumdar, D.~Nolte, F.~Quevedo, G.~Rajesh and R.-J. Zhang,
  \emph{{The Inflationary brane anti-brane universe}},
  \href{https://doi.org/10.1088/1126-6708/2001/07/047}{\emph{JHEP} {\bfseries
  07} (2001) 047}, [\href{https://arxiv.org/abs/hep-th/0105204}{{\ttfamily
  hep-th/0105204}}].

\bibitem{GarciaBellido:2001ky}
J.~Garcia-Bellido, R.~Rabadan and F.~Zamora, \emph{{Inflationary scenarios from
  branes at angles}},
  \href{https://doi.org/10.1088/1126-6708/2002/01/036}{\emph{JHEP} {\bfseries
  01} (2002) 036}, [\href{https://arxiv.org/abs/hep-th/0112147}{{\ttfamily
  hep-th/0112147}}].

\bibitem{Baumann:2007ah}
D.~Baumann, A.~Dymarsky, I.~R. Klebanov and L.~McAllister, \emph{{Towards an
  Explicit Model of D-brane Inflation}},
  \href{https://doi.org/10.1088/1475-7516/2008/01/024}{\emph{JCAP} {\bfseries
  0801} (2008) 024}, [\href{https://arxiv.org/abs/0706.0360}{{\ttfamily
  0706.0360}}].

\bibitem{Baumann:2014nda}
D.~Baumann and L.~McAllister, \emph{{Inflation and String Theory}}.
\newblock Cambridge Monographs on Mathematical Physics. Cambridge University
  Press, 2015,
  \href{https://doi.org/10.1017/CBO9781316105733}{10.1017/CBO9781316105733}.

\bibitem{Silverstein:2008sg}
E.~Silverstein and A.~Westphal, \emph{{Monodromy in the CMB: Gravity Waves and
  String Inflation}},
  \href{https://doi.org/10.1103/PhysRevD.78.106003}{\emph{Phys. Rev.}
  {\bfseries D78} (2008) 106003},
  [\href{https://arxiv.org/abs/0803.3085}{{\ttfamily 0803.3085}}].

\bibitem{McAllister:2008hb}
L.~McAllister, E.~Silverstein and A.~Westphal, \emph{{Gravity Waves and Linear
  Inflation from Axion Monodromy}},
  \href{https://doi.org/10.1103/PhysRevD.82.046003}{\emph{Phys. Rev.}
  {\bfseries D82} (2010) 046003},
  [\href{https://arxiv.org/abs/0808.0706}{{\ttfamily 0808.0706}}].

\bibitem{Dong:2010in}
X.~Dong, B.~Horn, E.~Silverstein and A.~Westphal, \emph{{Simple exercises to
  flatten your potential}},
  \href{https://doi.org/10.1103/PhysRevD.84.026011}{\emph{Phys. Rev.}
  {\bfseries D84} (2011) 026011},
  [\href{https://arxiv.org/abs/1011.4521}{{\ttfamily 1011.4521}}].

\bibitem{McAllister:2014mpa}
L.~McAllister, E.~Silverstein, A.~Westphal and T.~Wrase, \emph{{The Powers of
  Monodromy}}, \href{https://doi.org/10.1007/JHEP09(2014)123}{\emph{JHEP}
  {\bfseries 09} (2014) 123},
  [\href{https://arxiv.org/abs/1405.3652}{{\ttfamily 1405.3652}}].

\bibitem{Kallosh:2018nrk}
R.~Kallosh and T.~Wrase, \emph{{dS Supergravity from 10d}},
  \href{https://doi.org/10.1002/prop.201800071}{\emph{Fortsch. Phys.}
  {\bfseries 2018} (2018) 1800071},
  [\href{https://arxiv.org/abs/1808.09427}{{\ttfamily 1808.09427}}].

\bibitem{Blaback:2018hdo}
J.~Blaback, U.~Danielsson and G.~Dibitetto, \emph{{A new light on the darkest
  corner of the landscape}},
  \href{https://arxiv.org/abs/1810.11365}{{\ttfamily 1810.11365}}.

\bibitem{Polchinski:1996na}
J.~Polchinski, \emph{{Tasi lectures on D-branes}},  in \emph{{Fields, strings
  and duality. Proceedings, Summer School, Theoretical Advanced Study Institute
  in Elementary Particle Physics, TASI'96, Boulder, USA, June 2-28, 1996}},
  pp.~293--356, 1996, \href{https://arxiv.org/abs/hep-th/9611050}{{\ttfamily
  hep-th/9611050}}.

\bibitem{Bachas:1998rg}
C.~P. Bachas, \emph{{Lectures on D-branes}},  in \emph{{Duality and
  supersymmetric theories. Proceedings, Easter School, Newton Institute,
  Euroconference, Cambridge, UK, April 7-18, 1997}}, pp.~414--473, 1998,
  \href{https://arxiv.org/abs/hep-th/9806199}{{\ttfamily hep-th/9806199}}.

\bibitem{Kiritsis:1997hj}
E.~Kiritsis, \emph{{Introduction to superstring theory}}, vol.~B9 of
  \emph{Leuven notes in mathematical and theoretical physics}.
\newblock Leuven U. Press, Leuven, 1998.

\bibitem{Dasgupta:2002ew}
K.~Dasgupta, C.~Herdeiro, S.~Hirano and R.~Kallosh, \emph{{D3 / D7 inflationary
  model and M theory}},
  \href{https://doi.org/10.1103/PhysRevD.65.126002}{\emph{Phys. Rev.}
  {\bfseries D65} (2002) 126002},
  [\href{https://arxiv.org/abs/hep-th/0203019}{{\ttfamily hep-th/0203019}}].

\bibitem{Hsu:2003cy}
J.~P. Hsu, R.~Kallosh and S.~Prokushkin, \emph{{On brane inflation with volume
  stabilization}},
  \href{https://doi.org/10.1088/1475-7516/2003/12/009}{\emph{JCAP} {\bfseries
  0312} (2003) 009}, [\href{https://arxiv.org/abs/hep-th/0311077}{{\ttfamily
  hep-th/0311077}}].

\bibitem{Hsu:2004hi}
J.~P. Hsu and R.~Kallosh, \emph{{Volume stabilization and the origin of the
  inflaton shift symmetry in string theory}},
  \href{https://doi.org/10.1088/1126-6708/2004/04/042}{\emph{JHEP} {\bfseries
  04} (2004) 042}, [\href{https://arxiv.org/abs/hep-th/0402047}{{\ttfamily
  hep-th/0402047}}].

\bibitem{Kallosh:2007ig}
R.~Kallosh, \emph{{On inflation in string theory}},
  \href{https://doi.org/10.1007/978-3-540-74353-8_4}{\emph{Lect. Notes Phys.}
  {\bfseries 738} (2008) 119--156},
  [\href{https://arxiv.org/abs/hep-th/0702059}{{\ttfamily hep-th/0702059}}].

\bibitem{Haack:2008yb}
M.~Haack, R.~Kallosh, A.~Krause, A.~D. Linde, D.~Lust and M.~Zagermann,
  \emph{{Update of D3/D7-Brane Inflation on K3 x T**2/Z(2)}},
  \href{https://doi.org/10.1016/j.nuclphysb.2008.07.033}{\emph{Nucl. Phys.}
  {\bfseries B806} (2009) 103--177},
  [\href{https://arxiv.org/abs/0804.3961}{{\ttfamily 0804.3961}}].

\bibitem{Berg:2004ek}
M.~Berg, M.~Haack and B.~Kors, \emph{{Loop corrections to volume moduli and
  inflation in string theory}},
  \href{https://doi.org/10.1103/PhysRevD.71.026005}{\emph{Phys. Rev.}
  {\bfseries D71} (2005) 026005},
  [\href{https://arxiv.org/abs/hep-th/0404087}{{\ttfamily hep-th/0404087}}].

\bibitem{McAllister:2005mq}
L.~McAllister, \emph{{An Inflaton mass problem in string inflation from
  threshold corrections to volume stabilization}},
  \href{https://doi.org/10.1088/1475-7516/2006/02/010}{\emph{JCAP} {\bfseries
  0602} (2006) 010}, [\href{https://arxiv.org/abs/hep-th/0502001}{{\ttfamily
  hep-th/0502001}}].

\bibitem{Berg:2005ja}
M.~Berg, M.~Haack and B.~Kors, \emph{{String loop corrections to Kahler
  potentials in orientifolds}},
  \href{https://doi.org/10.1088/1126-6708/2005/11/030}{\emph{JHEP} {\bfseries
  11} (2005) 030}, [\href{https://arxiv.org/abs/hep-th/0508043}{{\ttfamily
  hep-th/0508043}}].

\bibitem{Witten:1996bn}
E.~Witten, \emph{{Nonperturbative superpotentials in string theory}},
  \href{https://doi.org/10.1016/0550-3213(96)00283-0}{\emph{Nucl. Phys.}
  {\bfseries B474} (1996) 343--360},
  [\href{https://arxiv.org/abs/hep-th/9604030}{{\ttfamily hep-th/9604030}}].

\bibitem{Gorlich:2004qm}
L.~Gorlich, S.~Kachru, P.~K. Tripathy and S.~P. Trivedi, \emph{{Gaugino
  condensation and nonperturbative superpotentials in flux compactifications}},
  \href{https://doi.org/10.1088/1126-6708/2004/12/074}{\emph{JHEP} {\bfseries
  12} (2004) 074}, [\href{https://arxiv.org/abs/hep-th/0407130}{{\ttfamily
  hep-th/0407130}}].

\bibitem{Bergshoeff:2005yp}
E.~Bergshoeff, R.~Kallosh, A.-K. Kashani-Poor, D.~Sorokin and A.~Tomasiello,
  \emph{{An Index for the Dirac operator on D3 branes with background fluxes}},
  \href{https://doi.org/10.1088/1126-6708/2005/10/102}{\emph{JHEP} {\bfseries
  10} (2005) 102}, [\href{https://arxiv.org/abs/hep-th/0507069}{{\ttfamily
  hep-th/0507069}}].

\bibitem{Ferrara:2014kva}
S.~Ferrara, R.~Kallosh and A.~Linde, \emph{{Cosmology with Nilpotent
  Superfields}}, \href{https://doi.org/10.1007/JHEP10(2014)143}{\emph{JHEP}
  {\bfseries 10} (2014) 143},
  [\href{https://arxiv.org/abs/1408.4096}{{\ttfamily 1408.4096}}].

\bibitem{Kallosh:2014wsa}
R.~Kallosh and T.~Wrase, \emph{{Emergence of Spontaneously Broken Supersymmetry
  on an Anti-D3-Brane in KKLT dS Vacua}},
  \href{https://doi.org/10.1007/JHEP12(2014)117}{\emph{JHEP} {\bfseries 12}
  (2014) 117}, [\href{https://arxiv.org/abs/1411.1121}{{\ttfamily 1411.1121}}].

\bibitem{Bergshoeff:2015jxa}
E.~A. Bergshoeff, K.~Dasgupta, R.~Kallosh, A.~Van~Proeyen and T.~Wrase,
  \emph{{$ \overline{\mathrm{D}3} $ and dS}},
  \href{https://doi.org/10.1007/JHEP05(2015)058}{\emph{JHEP} {\bfseries 05}
  (2015) 058}, [\href{https://arxiv.org/abs/1502.07627}{{\ttfamily
  1502.07627}}].

\bibitem{Kallosh:2015nia}
R.~Kallosh, F.~Quevedo and A.~M. Uranga, \emph{{String Theory Realizations of
  the Nilpotent Goldstino}},
  \href{https://doi.org/10.1007/JHEP12(2015)039}{\emph{JHEP} {\bfseries 12}
  (2015) 039}, [\href{https://arxiv.org/abs/1507.07556}{{\ttfamily
  1507.07556}}].

\bibitem{Bergshoeff:2015tra}
E.~A. Bergshoeff, D.~Z. Freedman, R.~Kallosh and A.~Van~Proeyen, \emph{{Pure de
  Sitter Supergravity}}, \href{https://doi.org/10.1103/PhysRevD.93.069901,
  10.1103/PhysRevD.92.085040}{\emph{Phys. Rev.} {\bfseries D92} (2015) 085040},
  [\href{https://arxiv.org/abs/1507.08264}{{\ttfamily 1507.08264}}].

\bibitem{Hasegawa:2015bza}
F.~Hasegawa and Y.~Yamada, \emph{{Component action of nilpotent multiplet
  coupled to matter in 4 dimensional $ \mathcal{N}=1 $ supergravity}},
  \href{https://doi.org/10.1007/JHEP10(2015)106}{\emph{JHEP} {\bfseries 10}
  (2015) 106}, [\href{https://arxiv.org/abs/1507.08619}{{\ttfamily
  1507.08619}}].

\bibitem{Volkov:1972jx}
D.~V. Volkov and V.~P. Akulov, \emph{{Possible universal neutrino
  interaction}}, {\emph{JETP Lett.} {\bfseries 16} (1972) 438--440}.

\bibitem{McDonough:2016der}
E.~McDonough and M.~Scalisi, \emph{{Inflation from Nilpotent K\"ahler
  Corrections}},
  \href{https://doi.org/10.1088/1475-7516/2016/11/028}{\emph{JCAP} {\bfseries
  1611} (2016) 028}, [\href{https://arxiv.org/abs/1609.00364}{{\ttfamily
  1609.00364}}].

\bibitem{Kallosh:2017wnt}
R.~Kallosh, A.~Linde, D.~Roest and Y.~Yamada, \emph{{$ \overline{D3} $ induced
  geometric inflation}},
  \href{https://doi.org/10.1007/JHEP07(2017)057}{\emph{JHEP} {\bfseries 07}
  (2017) 057}, [\href{https://arxiv.org/abs/1705.09247}{{\ttfamily
  1705.09247}}].

\bibitem{Kachru:2003aw}
S.~Kachru, R.~Kallosh, A.~D. Linde and S.~P. Trivedi, \emph{{De Sitter vacua in
  string theory}},
  \href{https://doi.org/10.1103/PhysRevD.68.046005}{\emph{Phys. Rev.}
  {\bfseries D68} (2003) 046005},
  [\href{https://arxiv.org/abs/hep-th/0301240}{{\ttfamily hep-th/0301240}}].

\bibitem{Kallosh:2011qk}
R.~Kallosh, A.~Linde, K.~A. Olive and T.~Rube, \emph{{Chaotic inflation and
  supersymmetry breaking}},
  \href{https://doi.org/10.1103/PhysRevD.84.083519}{\emph{Phys. Rev.}
  {\bfseries D84} (2011) 083519},
  [\href{https://arxiv.org/abs/1106.6025}{{\ttfamily 1106.6025}}].

\bibitem{Linde:2011ja}
A.~Linde, Y.~Mambrini and K.~A. Olive, \emph{{Supersymmetry Breaking due to
  Moduli Stabilization in String Theory}},
  \href{https://doi.org/10.1103/PhysRevD.85.066005}{\emph{Phys. Rev.}
  {\bfseries D85} (2012) 066005},
  [\href{https://arxiv.org/abs/1111.1465}{{\ttfamily 1111.1465}}].

\bibitem{Dudas:2012wi}
E.~Dudas, A.~Linde, Y.~Mambrini, A.~Mustafayev and K.~A. Olive, \emph{{Strong
  moduli stabilization and phenomenology}},
  \href{https://doi.org/10.1140/epjc/s10052-012-2268-7}{\emph{Eur. Phys. J.}
  {\bfseries C73} (2013) 2268},
  [\href{https://arxiv.org/abs/1209.0499}{{\ttfamily 1209.0499}}].

\bibitem{Kallosh:2004yh}
R.~Kallosh and A.~D. Linde, \emph{{Landscape, the scale of SUSY breaking, and
  inflation}}, \href{https://doi.org/10.1088/1126-6708/2004/12/004}{\emph{JHEP}
  {\bfseries 12} (2004) 004},
  [\href{https://arxiv.org/abs/hep-th/0411011}{{\ttfamily hep-th/0411011}}].

\bibitem{BlancoPillado:2005fn}
J.~J. Blanco-Pillado, R.~Kallosh and A.~D. Linde, \emph{{Supersymmetry and
  stability of flux vacua}},
  \href{https://doi.org/10.1088/1126-6708/2006/05/053}{\emph{JHEP} {\bfseries
  05} (2006) 053}, [\href{https://arxiv.org/abs/hep-th/0511042}{{\ttfamily
  hep-th/0511042}}].

\bibitem{Kallosh:2014oja}
R.~Kallosh, A.~Linde, B.~Vercnocke and T.~Wrase, \emph{{Analytic Classes of
  Metastable de Sitter Vacua}},
  \href{https://doi.org/10.1007/JHEP10(2014)011}{\emph{JHEP} {\bfseries 10}
  (2014) 011}, [\href{https://arxiv.org/abs/1406.4866}{{\ttfamily 1406.4866}}].

\bibitem{Hasegawa:2017nks}
F.~Hasegawa, K.~Nakayama, T.~Terada and Y.~Yamada, \emph{{Gravitino problem in
  inflation driven by inflaton-polonyi K\"ahler coupling}},
  \href{https://doi.org/10.1016/j.physletb.2017.12.038}{\emph{Phys. Lett.}
  {\bfseries B777} (2018) 270--274},
  [\href{https://arxiv.org/abs/1709.01246}{{\ttfamily 1709.01246}}].

\bibitem{Spergel:2003cb}
{\scshape WMAP} collaboration, D.~N. Spergel et~al., \emph{{First year
  Wilkinson Microwave Anisotropy Probe (WMAP) observations: Determination of
  cosmological parameters}},
  \href{https://doi.org/10.1086/377226}{\emph{Astrophys. J. Suppl.} {\bfseries
  148} (2003) 175--194},
  [\href{https://arxiv.org/abs/astro-ph/0302209}{{\ttfamily
  astro-ph/0302209}}].

\bibitem{Hinshaw:2012aka}
{\scshape WMAP} collaboration, G.~Hinshaw et~al., \emph{{Nine-Year Wilkinson
  Microwave Anisotropy Probe (WMAP) Observations: Cosmological Parameter
  Results}}, \href{https://doi.org/10.1088/0067-0049/208/2/19}{\emph{Astrophys.
  J. Suppl.} {\bfseries 208} (2013) 19},
  [\href{https://arxiv.org/abs/1212.5226}{{\ttfamily 1212.5226}}].

\bibitem{Ade:2013zuv}
{\scshape Planck} collaboration, P.~A.~R. Ade et~al., \emph{{Planck 2013
  results. XVI. Cosmological parameters}},
  \href{https://doi.org/10.1051/0004-6361/201321591}{\emph{Astron. Astrophys.}
  {\bfseries 571} (2014) A16},
  [\href{https://arxiv.org/abs/1303.5076}{{\ttfamily 1303.5076}}].

\bibitem{Aghanim:2018eyx}
{\scshape Planck} collaboration, N.~Aghanim et~al., \emph{{Planck 2018 results.
  VI. Cosmological parameters}},
  \href{https://arxiv.org/abs/1807.06209}{{\ttfamily 1807.06209}}.

\end{thebibliography}\endgroup

\end{document}